\documentclass[preprint,superscriptaddress, secnumarabic,amssymb, nobibnotes, aps, prd, showpacs]{revtex4-1}

\usepackage{hyperref}
\usepackage{graphics}
\usepackage{graphicx}
\usepackage{float}
\usepackage{longtable}
\begin{document}

\title{Measurements of $W$ and $Z/\gamma^*$ cross sections and their ratios in $p+p$ collisions at RHIC}%

\affiliation{Abilene Christian University, Abilene, Texas   79699}
\affiliation{AGH University of Science and Technology, FPACS, Cracow 30-059, Poland}
\affiliation{Alikhanov Institute for Theoretical and Experimental Physics NRC "Kurchatov Institute", Moscow 117218, Russia}
\affiliation{Argonne National Laboratory, Argonne, Illinois 60439}
\affiliation{American University of Cairo, New Cairo 11835, New Cairo, Egypt}
\affiliation{Brookhaven National Laboratory, Upton, New York 11973}
\affiliation{University of California, Berkeley, California 94720}
\affiliation{University of California, Davis, California 95616}
\affiliation{University of California, Los Angeles, California 90095}
\affiliation{University of California, Riverside, California 92521}
\affiliation{Central China Normal University, Wuhan, Hubei 430079 }
\affiliation{University of Illinois at Chicago, Chicago, Illinois 60607}
\affiliation{Creighton University, Omaha, Nebraska 68178}
\affiliation{Czech Technical University in Prague, FNSPE, Prague 115 19, Czech Republic}
\affiliation{Technische Universit\"at Darmstadt, Darmstadt 64289, Germany}
\affiliation{ELTE E\"otv\"os Lor\'and University, Budapest, Hungary H-1117}
\affiliation{Frankfurt Institute for Advanced Studies FIAS, Frankfurt 60438, Germany}
\affiliation{Fudan University, Shanghai, 200433 }
\affiliation{University of Heidelberg, Heidelberg 69120, Germany }
\affiliation{University of Houston, Houston, Texas 77204}
\affiliation{Huzhou University, Huzhou, Zhejiang  313000}
\affiliation{Indian Institute of Science Education and Research (IISER), Berhampur 760010 , India}
\affiliation{Indian Institute of Science Education and Research (IISER) Tirupati, Tirupati 517507, India}
\affiliation{Indian Institute Technology, Patna, Bihar 801106, India}
\affiliation{Indiana University, Bloomington, Indiana 47408}
\affiliation{Institute of Modern Physics, Chinese Academy of Sciences, Lanzhou, Gansu 730000 }
\affiliation{University of Jammu, Jammu 180001, India}
\affiliation{Joint Institute for Nuclear Research, Dubna 141 980, Russia}
\affiliation{Kent State University, Kent, Ohio 44242}
\affiliation{University of Kentucky, Lexington, Kentucky 40506-0055}
\affiliation{Lawrence Berkeley National Laboratory, Berkeley, California 94720}
\affiliation{Lehigh University, Bethlehem, Pennsylvania 18015}
\affiliation{Max-Planck-Institut f\"ur Physik, Munich 80805, Germany}
\affiliation{Michigan State University, East Lansing, Michigan 48824}
\affiliation{National Research Nuclear University MEPhI, Moscow 115409, Russia}
\affiliation{National Institute of Science Education and Research, HBNI, Jatni 752050, India}
\affiliation{National Cheng Kung University, Tainan 70101 }
\affiliation{Nuclear Physics Institute of the CAS, Rez 250 68, Czech Republic}
\affiliation{Ohio State University, Columbus, Ohio 43210}
\affiliation{Institute of Nuclear Physics PAN, Cracow 31-342, Poland}
\affiliation{Panjab University, Chandigarh 160014, India}
\affiliation{Pennsylvania State University, University Park, Pennsylvania 16802}
\affiliation{NRC "Kurchatov Institute", Institute of High Energy Physics, Protvino 142281, Russia}
\affiliation{Purdue University, West Lafayette, Indiana 47907}
\affiliation{Rice University, Houston, Texas 77251}
\affiliation{Rutgers University, Piscataway, New Jersey 08854}
\affiliation{Universidade de S\~ao Paulo, S\~ao Paulo, Brazil 05314-970}
\affiliation{University of Science and Technology of China, Hefei, Anhui 230026}
\affiliation{Shandong University, Qingdao, Shandong 266237}
\affiliation{Shanghai Institute of Applied Physics, Chinese Academy of Sciences, Shanghai 201800}
\affiliation{Southern Connecticut State University, New Haven, Connecticut 06515}
\affiliation{State University of New York, Stony Brook, New York 11794}
\affiliation{Instituto de Alta Investigaci\'on, Universidad de Tarapac\'a, Arica 1000000, Chile}
\affiliation{Temple University, Philadelphia, Pennsylvania 19122}
\affiliation{Texas A\&M University, College Station, Texas 77843}
\affiliation{University of Texas, Austin, Texas 78712}
\affiliation{Tsinghua University, Beijing 100084}
\affiliation{University of Tsukuba, Tsukuba, Ibaraki 305-8571, Japan}
\affiliation{United States Naval Academy, Annapolis, Maryland 21402}
\affiliation{Valparaiso University, Valparaiso, Indiana 46383}
\affiliation{Variable Energy Cyclotron Centre, Kolkata 700064, India}
\affiliation{Warsaw University of Technology, Warsaw 00-661, Poland}
\affiliation{Wayne State University, Detroit, Michigan 48201}
\affiliation{Yale University, New Haven, Connecticut 06520}

\author{J.~Adam}\affiliation{Brookhaven National Laboratory, Upton, New York 11973}
\author{L.~Adamczyk}\affiliation{AGH University of Science and Technology, FPACS, Cracow 30-059, Poland}
\author{J.~R.~Adams}\affiliation{Ohio State University, Columbus, Ohio 43210}
\author{J.~K.~Adkins}\affiliation{University of Kentucky, Lexington, Kentucky 40506-0055}
\author{G.~Agakishiev}\affiliation{Joint Institute for Nuclear Research, Dubna 141 980, Russia}
\author{M.~M.~Aggarwal}\affiliation{Panjab University, Chandigarh 160014, India}
\author{Z.~Ahammed}\affiliation{Variable Energy Cyclotron Centre, Kolkata 700064, India}
\author{I.~Alekseev}\affiliation{Alikhanov Institute for Theoretical and Experimental Physics NRC "Kurchatov Institute", Moscow 117218, Russia}\affiliation{National Research Nuclear University MEPhI, Moscow 115409, Russia}
\author{D.~M.~Anderson}\affiliation{Texas A\&M University, College Station, Texas 77843}
\author{A.~Aparin}\affiliation{Joint Institute for Nuclear Research, Dubna 141 980, Russia}
% ...

\collaboration{STAR Collaboration}\noaffiliation

\date{November 11, 2020}%

\begin{abstract}
We report on the $W$ and $Z/\gamma^*$ differential and total cross sections as well as the $W^+$/$W^-$ and $(W^+ + W^-)$/$(Z/\gamma^*)$ cross-section ratios measured by the STAR experiment at RHIC in $p+p$ collisions at $\sqrt{s} = 500$ GeV and $510$ GeV. The cross sections and their ratios are sensitive to quark and antiquark parton distribution functions. In particular, at leading order, the $W$ cross-section ratio is sensitive to the $\bar{d}/\bar{u}$ ratio. These measurements were taken at high $Q^2 \sim M_W^2,M_Z^2$ and can serve as input into global analyses to provide constraints on the sea quark distributions. The results presented here combine three STAR data sets from 2011, 2012, and 2013, accumulating an integrated luminosity of 350 pb$^{-1}$. We also assess the expected impact that our $W^+/W^-$ cross-section ratios will have on various quark distributions, and find sensitivity to the $\bar{u}-\bar{d}$ and $\bar{d}/\bar{u}$ distributions. 
\end{abstract}

%\keywords{W production; Z production}

\maketitle
%\tableofcontents

\section{Introduction}
\label{sec:intro}
Since the discovery of the $W$ and $Z$ bosons by the UA1~\cite{ARNISON1983103,1983398,Albajar:1987yz,Albajar:1988ka} and UA2~\cite{BANNER1983476,Bagnaia:146503,Alitti:1990gj,Alitti:1991dm} experiments in proton-antiproton collisions at the CERN S$p\bar{p}$S facility, a significant amount of work has been done measuring the properties of the bosons using a variety of collision systems. These probes range from additional proton-antiproton collision measurements by CDF~\cite{Abe:1995bm,Abulencia:2005ix,Aaltonen:2012fi,Aaltonen:2012bp} and D0~\cite{Abachi:1995xc,Abbott:1999gn,Abbott:1999tt,Abazov:2007ac,Abazov:2012bv} at the Fermilab Tevatron, to measurements based on electron-positron collisions by the ALEPH, DELPHI, L3, and OPAL experiments performed at LEP~\cite{Schael:2013ita,2006257,Ward:2005pn}. More recent measurements from ATLAS~\cite{Aad:2010yt,Aad:2016naf,Aaboud:2016btc,Aaboud:2018nic} and CMS~\cite{Khachatryan:2010xn,CMS:2011aa,CMS:2016mtd,CMS:2015ois} at the LHC, and PHENIX~\cite{Adare:2010xa,Adare:2018csm} and STAR~\cite{STAR:2011aa} at the Relativistic Heavy Ion Collider (RHIC) use proton-proton collisions to investigate the properties of the $W$ and $Z$ bosons. Additionally, both the PHENIX and STAR experiments have used polarized proton collisions to study the $W$ and $Z$ boson spin asymmetries~\cite{Adare:2018csm,Adare:2015gsd,Aggarwal:2010vc,Adamczyk:2014xyw,Adamczyk:2015gyk,Adam:2018bam}. The current study of inclusive $W$ and $Z$ boson production benefits from these previous experiments. Modern measurements not only serve as an excellent benchmark for Standard Model testing, but also as a means by which to constrain Parton Distribution Functions (PDFs) of the proton. 

One particular parton distribution of interest is the $\bar{d}/\bar{u}$ ratio near the valence region ($x \approx 0.3$). While the PDFs that characterize the valence quarks in the proton are well determined from deep inelastic scattering experiments, the antiquarks are less known. Over the years, Drell-Yan experiments~\cite{Baldit:1994jk,Towell:2001nh,Nagai:2018scg,Nakahara:2011zz} have probed the $\bar{d}/\bar{u}$ distribution in the proton. The NuSea experiment found evidence of a larger-than-expected $\bar{d}/\bar{u}$ flavor asymmetry, especially as $x$, the fraction of the proton momentum carried by the struck parton, exceeds $x\approx 0.2$~\cite{Towell:2001nh}. While the SeaQuest experiment (still under analysis at the time of this writing~\cite{Nagai:2018scg,Nakahara:2011zz}) will push the measurement to larger $x$ and improve on statistics compared to the previous NuSea measurement, the STAR experiment at RHIC is able to provide new and complementary information about the $\bar{d}/\bar{u}$ distribution, from a different reaction channel, $W$ production, at a large momentum scale, $Q^2 = M^2_W$.  

RHIC can collide protons up to $\sqrt{s}=510$ GeV. $W^\pm$ bosons at RHIC are produced through $u+\bar{d}\;(d+\bar{u})$ fusion, which allows observables to have sensitivity to the sea quark distributions. The $W^+$/$W^-$ cross section ratio is sensitive to the $\bar{d}/\bar{u}$ distribution, as can be seen from its leading order contribution~\cite{BOURRELY1994329}
\begin{equation}\label{eq:RW-Theory}
\frac{\sigma_{W^+}}{\sigma_{W^-}} \approx \frac{u(x_1)\bar{d}(x_2)+ u(x_2)\bar{d}(x_1)}{d(x_1)\bar{u}(x_2) + d(x_2)\bar{u}(x_1)},
\end{equation}   
where $x_1$ and $x_2$ are the fractions of the proton momenta carried by the scattering partons. 
Additionally, ATLAS has recently used their measured $(W^+ + W^-)/Z$ cross-section ratio to investigate the strange quark content of the proton~\cite{Aaboud:2016btc}, where an enhancement of the proton strange quark contribution is seen. Furthermore, measurements of differential $W$ and $Z$ cross sections have been used to provide further constraints for PDF extractions~\cite{Aaboud:2016btc,Aad:2012sb}. These quantities measured at STAR serve as complementary measurements to their LHC counterparts. They probe a higher $x$ region due to the lower center of mass energy of the proton collisions.

We report on the measurements of the differential and total $W$ and $Z$ cross sections, as well as the $W^+/W^-$ and $(W^+ + W^-)$/$(Z/\gamma^*)$ cross-section ratios made by the STAR experiment at RHIC during the 2011, 2012, and 2013 $p+p$ running periods at $\sqrt{s} = 500$ GeV (2011 data set) and $510$ GeV (2012 and 2013 data sets), accumulating a total integrated luminosity of $350$ pb$^{-1}$. A summary of these data sets, including their center of mass energies and integrated luminosities, is listed in Table~\ref{tbl:dataset}. These measurements are derived from studies of the $W^{+(-)}\rightarrow e^{+(-)} + \nu(\bar{\nu})$ and $Z/\gamma^*\rightarrow e^+e^-$ decay channels for outgoing leptons. This expands on previous STAR results based on the RHIC 2009 $p+p$ data set~\cite{STAR:2011aa}, not only by adding more statistics, but also in several other areas. First, in addition to the total $W$ and $Z$ cross sections, we have measured the differential cross sections $d\sigma_{W^\pm}/d\eta_{e^\pm}$ and $d\sigma_{Z/\gamma^*}/dy_Z$ as functions of $e^\pm$ pseudorapidity, $\eta_{e^\pm}$, and $Z$ boson rapidity, $y_Z$, respectively. Second, a measurement of the lepton pseudorapidity dependence of the $W^+$/$W^-$ cross-section ratio between $-1.0 \le \eta \le 1.5$ was made. Finally, the $(W^+ + W^-)$/$(Z/\gamma^*)$ cross-section ratio was measured. These measurements make use of the same apparatus and techniques described in previous STAR $W$ and $Z$ publications~\cite{STAR:2011aa,Aggarwal:2010vc,Adamczyk:2014xyw,Adamczyk:2015gyk,Adam:2018bam}.  

Our results are organized into eight additional sections. Section~\ref{sec:expsetup} provides a brief overview of the STAR subsystems used in this analysis, while Sec.~\ref{sec:data} describes the data and simulation samples that were used. The details regarding the extraction of the $W$ and $Z/\gamma^*$  signals from the data and the procedures used to estimate the background contributions are discussed in Secs.~\ref{sec:recon} and \ref{sec:background}. In Sec.~\ref{sec:eff} we report on the electron and positron detection efficiencies. The differential and total cross section results are presented in Sec.~\ref{sec:xsec}, while the $W^+$/$W^-$ and $(W^+ + W^-)$/$Z$ cross-section ratios are shown in Sec.~\ref{sec:ratios}. Finally, Sec.~\ref{sec:summary} presents a summary of the measurements. Throughout the remainder of the paper we will be using~``$Z/\gamma^*$" and ``$Z$" interchangeably.

\section{Experimental Setup}
\label{sec:expsetup}
The Solenoidal Tracker At RHIC (STAR) detector~\cite{ACKERMANN2003624} and its subsystems have been thoroughly described in similar STAR analyses~\cite{STAR:2011aa,Aggarwal:2010vc,Adamczyk:2014xyw,Adamczyk:2015gyk,Adam:2018bam}. The presented analysis utilizes several subsystems of the STAR detector. Charged particle tracking, including momentum reconstruction and charge sign determination, is provided by the Time Projection Chamber (TPC)~\cite{ANDERSON2003659} in combination with a 0.5 T magnetic field. The TPC lies between 50 and 200 cm from the beam axis and covers pseudorapidities $|\eta|<1.3$ and the full azimuthal angle, $0 < \phi < 2\pi$.

Surrounding the TPC is the Barrel Electromagnetic Calorimeter (BEMC)~\cite{BEDDO2003725}, which is a lead-scintillator sampling calorimeter. The BEMC is segmented into 4800 optically isolated towers covering the full azimuthal angle for pseudorapidities $|\eta|<1$, referred to in this paper as the mid-pseudorapidity region.
 
A second lead-scintillator based calorimeter is located at one end of the STAR TPC, the Endcap Electromagnetic Calorimeter (EEMC)~\cite{ALLGOWER2003740}. The EEMC consists of 720 towers extending the particle energy deposition measurements to a pseudorapidity of $1.1 < \eta < 2.0$, referred to as the intermediate pseudorapidity region, while maintaining full azimuthal coverage. Included within the EEMC is the EEMC Shower Maximum Detector (ESMD)~\cite{ALLGOWER2003740}, which is used to discriminate amongst isolated electron or positron (signal) events and wider showers typically seen from jet-like events (background). This discrimination is determined by measuring the transverse profile of the electromagnetic shower. The ESMD consists of scintillator strips organized into orthogonal $u$ and $v$ planes. 

Finally, the Zero Degree Calorimeter (ZDC)~\cite{ACKERMANN2003624} is used to determine and monitor the luminosity.
\section{Data and Simulation}
\label{sec:data}
We present results based on measuremnts made in the mid- ($|\eta_e| < 1.0$) and intermediate pseudorapidity ( $1.0 < \eta_e < 1.5$ ) regions. The mid-pseudorapidity region measurements combined data that were recorded during the 2011, 2012, and 2013 STAR $p+p$ running periods (Table~\ref{tbl:dataset}). Due to insufficient statistics collected in the intermediate pseudorapidity range during the 2011 running period, measurements made in this region only combined the data taken during the 2012 and 2013 running periods.

Before combining the mid-pseudorapidity 2011 data set (taken at $\sqrt{s}$ = 500 GeV) with the mid-pseudorapidity 2012 and 2013 data sets (taken at $\sqrt{s}$ = 510 GeV), we studied how the $W$ and $Z$ fiducial cross sections changed between the two center of mass energies. The study was performed using the FEWZ~\cite{PhysRevD.86.094034} theory code with the CT14 PDF set~\cite{PhysRevD.93.033006}, and calculated a 4.7\%, 5.4\%, and 6\% larger $W^+$, $W^-$, and $Z$ cross section, respectively, for the higher center of mass energy. To account for these differences, we scaled our measured 2011 $W$ and $Z$ fiducial cross sections by the ratio of the cross sections at $\sqrt{s} = 510$ GeV to the cross sections at $\sqrt{s} = 500$ GeV, computed from the FEWZ-CT14 study, for each of our lepton pseudorapidity and $Z$ rapidity data bins. These corrections ($\approx 5-6\% $) have a small effect overall since the 2011 data set only makes up roughly 7\% of the combined data set.  

The integrated luminosity for each data set is needed to normalize the measured cross sections and was determined using the standard RHIC Van Der Meer Scan technique~\cite{vanderMeer:296752,STAR:2011aa,DreesCAD:2013}. Based on this technique we have estimated an overall uncertainty of 9\% for the integrated luminosity.

$W^{+(-)}$ and $Z/\gamma^*$ bosons were detected via the leptonic decay channels $W^{+(-)}\rightarrow e^{+(-)} + \nu(\bar{\nu})$ and $Z/\gamma^*\rightarrow e^+ + e^-$.
Events that pass a calorimeter trigger, which required a transverse energy, $E_T$, covering a region  of $\approx 0.1\times 0.1$ in $\Delta \phi \times \Delta \eta$, to be greater than 12 (10) GeV in the BEMC (EEMC), constitute our initial $W/Z$ decay candidate sample. This sample of events is later refined by applying additional selection criteria, as discussed in Sec.~\ref{sec:recon}.
%The calorimeters were used to select candidates online using a two-level triggering system. The hardware level-0 trigger acceted events containing a tower with transverse energy,$E_T$ greater than $\Rightarrow 7.3$ GeV. A dedicated software trigger algorithm then formed 2x2 tower clusters, covering a region of $\approx 0.1\times 0.1$ in $\Delta \phi \times \Delta \eta$, and requireing that $\Rightarrow E_T > 12$ and $10$ GeV for the BEMC and EEMC, respectivly. 
\begin{table}[tb]
\caption{\label{tbl:dataset} Summary of data sets used in this analysis.}
\begin{ruledtabular}
\begin{tabular}{ccc}
Data Sample & $\sqrt{s}$ (GeV)& $\mathcal{L}$ (pb$^{-1}$)\\
\hline
2011 & 500 &  $25 \pm 2$\\
2012 & 510 &  $75 \pm 7$\\
2013 & 510 &  $250\pm 22$\\ 
\end{tabular}
\end{ruledtabular}
\end{table}

In order to determine detector efficiencies and estimate background contributions from electroweak processes, Monte Carlo (MC) samples for $Z/\gamma^* \rightarrow e^+e^-$, $W \rightarrow e\nu$, and $W \rightarrow \tau \nu$ were generated. All samples were produced using PYTHIA 6.4.28~\cite{Sj_strand_2006} and the~``Perugia 0" tune~\cite{PhysRevD.82.074018}. The event distributions were then passed through a GEANT 3~\cite{Brun:1994aa} model of the STAR detector, after which they were embedded into STAR zero-bias data to account for pile-up tracks in the TPC volume. The pile-up tracks can be caused by another collision from the same bunch crossing as the triggered event, or a collision that occurred in an earlier or later bunch crossing. The zero-bias events were obtained during bunch crossings that were recorded with no cuts applied. Finally, the MC samples were weighted with the integrated luminosity of the respective STAR data set. The same reconstruction and analysis algorithms were used on both the MC and data samples.
\section{$W$ and $Z/\gamma^*$ Reconstruction}
\label{sec:recon}
$W$ and $Z/\gamma^*$ candidate events were identified and reconstructed using well-established selection cuts used in past STAR measurements~\cite{STAR:2011aa,Aggarwal:2010vc,Adamczyk:2014xyw,Adamczyk:2015gyk,Adam:2018bam}. Candidate events that triggered the electromagnetic calorimeters are required to have their collision vertex along the beam axis within 100 cm of the center of STAR. The vertex was reconstructed using tracks measured in the TPC. The reconstructed vertices had a distribution along the beam axis that was roughly Gaussian with an RMS width of about 40 cm.

In addition to the conditions discussed above, a candidate electron or positron track at mid-pseudorapidity (intermediate pseudorapidity) with an associated reconstructed vertex was also required to have transverse momentum, $p_T$, larger than 10 (7) GeV. To help ensure that the track and its charge sign are well reconstructed, and to remove pile-up tracks which may have accidentlly been associated with a vertex, we implemented several TPC related requirements. First, we required that the reconstructed track has at least 15 (5) TPC hit points. Secondly, the number of hit points used in the track fitting needed to be more than 51\% of the possible hit points. Finally, in the mid-pseudorapidity range we required that the first TPC hit point has a radius (with respect to the beam axis) less than 90 cm, while the last TPC hit point had a radius greater than 160 cm. A modified cut was applied to tracks in the intermediate pseudorapidity region, where the first TPC hit point was required to have a radius smaller than 120 cm.    
  
The transverse energy of the $e^\pm$ decay candidates, $E^e_T$, was determined from the largest transverse energy $2\times2$ calorimeter cluster that contains the triggered tower. We required that this energy be greater than 16 (20) GeV for the BEMC (EEMC) and that the candidate's track projected to within 7 (10) cm of the cluster center.  

\subsection{Electron and Positron Isolation Cuts}\label{sec:isolationcuts}
Electrons and positrons originating from $W$ and $Z$ decays should be relatively isolated from other particles in $\eta-\phi$ space, resulting in isolated transverse energy deposition in the BEMC and EEMC calorimeters. Jet-like events can be reduced by employing several isolation cuts. The first cut requires the ratio of the $e^\pm$ candidate's $E^e_T$  and the total $E_T$ from a $4\times 4$ BEMC (EEMC) cluster surrounding the $e^\pm$ candidate $2\times 2$ tower cluster to be greater than 96\% (97\%). For the second cut, the ratio of the $e^\pm$ candidate's  $E^e_T$ to the transverse energy, $E^{\Delta R<0.7}_T$, within a cone of radius $\Delta R = \sqrt{\eta^2 + \phi^2} < 0.7$ around the candidate track was required to be greater than 82\% (88\%). The transverse energy $E^{\Delta R<0.7}_T$ was determined by summing the BEMC and EEMC $E_T$ and the TPC track $p_T$ within the cone. The $e^\pm$ candidate track was excluded from the sum of TPC track $p_T$ to avoid double-counting in $E^{\Delta R<0.7}_T$. The final isolation cut only applies to the EEMC and in particular the ESMD. The ESMD can be used to discriminate between isolated $e^\pm$, which could come from $W$ and $Z$ decays, and QCD/jet-like events by measuring the transverse profile of the electromagnetic shower in the two ESMD layers. The transverse profile of the electromagnetic shower resulting from isolated $e^\pm$ will be narrower than the profiles produced from QCD and jet-like backgrounds. TPC tracks were extrapolated to the ESMD, where the central strip in each direction was defined as the nearest strip pointed to by the track. A ratio, $R_{ESMD}$, was formed with a numerator equal to the total energy deposited in the ESMD strips that were within 1.5 cm of the central strips, and a denominator equal to the total energy deposited in the strips that were within 10 cm of the central strips. For this analysis, we required this ratio to be larger than 70\%.

\subsection{$W^\pm$ Candidate Event Selection}\label{sec:wcandidate}
Differences in the event topologies between leptonic $W$ decays and QCD or $Z$ decays can be used to select $W\rightarrow e\nu$ candidate events. A $\vec{p}\,^{bal}_T$ vector can be constructed which is the vector sum of the decay $e^\pm$ transverse momentum, $\vec{p}\,^e_{T}$, plus the sum of $\vec{p}_T$ vectors for jets reconstructed outside of a cone radius $\Delta R = 0.7$. Using towers with $E_T > 0.2$ GeV and tracks with $p_T > 0.2$ GeV, the jets were reconstructed using the anti-$k_T$ algorithm~\cite{Cacciari_2008} in which the resolution parameter was set to $0.6$. Reconstructed jets were required to have $p_T > 3.5$ GeV. $W$ candidates will possess a large missing transverse momentum, due to the undetected neutrino, which leads to a large imbalance when computing $\vec{p}\,^{bal}_T$. In contrast, $Z\rightarrow e^+e^-$ and QCD backgrounds, such as dijets, do not produce such a large $\vec{p}\,^{bal}_T$. Therefore, using the $\vec{p}\,^{bal}_T$ vector we define a scalar signed-$p_T$ balance quantity as $\left(\vec{p}\,^e_T \cdot \vec{p}\,^{bal}_T \right)/|\vec{p}\,^e_T|$ and require it to be larger than 16 (20) GeV for $e^\pm$ candidates detected in the BEMC (EEMC). In addition to the signed-$p_T$ balance cut, the total transverse energy opposite the candidate electron or positron in the BEMC ($|\Delta\phi - \pi| < 0.7$) was required not to exceed $11$ GeV. This further helped to remove QCD dijet background events where a sizable fraction of the energy of one of the jets was not observed due to detector effects. Due to the effectiveness of the $R_{ESMD}$ cut, the cut on the transverse energy opposite of the candidate electron or positron was not needed in the EEMC.
\begin{figure}[tbh]
\center
\includegraphics[width = 1\columnwidth,keepaspectratio]{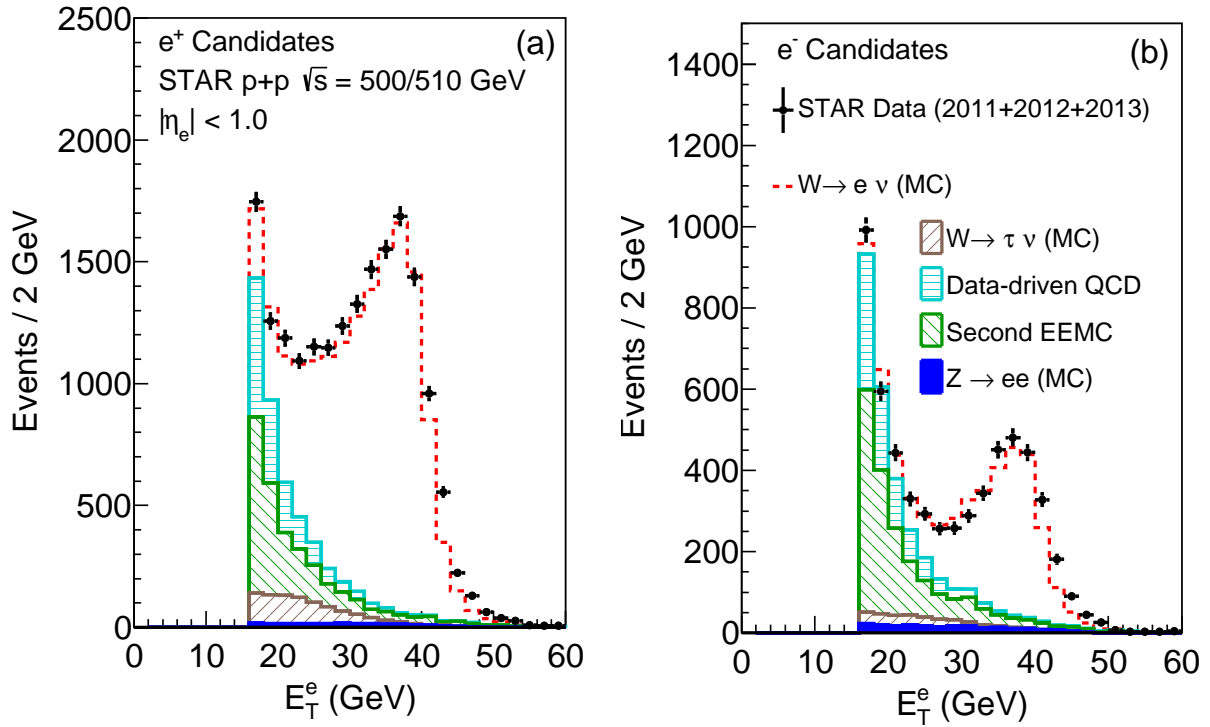}
\caption{\label{fig:BEMCbkgd} Signal and background $E^e_T$ distributions for positron (a) and electron (b) candidates in the BEMC. The background contributions are shown as stacked histograms, where the solid blue and brown diagonal histograms correspond to the electroweak residual backgrounds from $Z\rightarrow ee$ and $W\rightarrow \tau\nu$ decay channels, respectively. The vertical cyan and diagonal green histograms correspond to the residual QCD contributions estimated from the data driven and second EEMC methods, respectively. The red dashed histogram shows the $W\rightarrow e\nu$ signal along with all estimated background contributions and is compared to the data, the black markers. The vertical error bar on the data represents the statistical uncertainty and the horizontal bar shows the bin width.} 
\end{figure}
\begin{figure}[tbh]
\center
\includegraphics[width = 1\columnwidth,keepaspectratio]{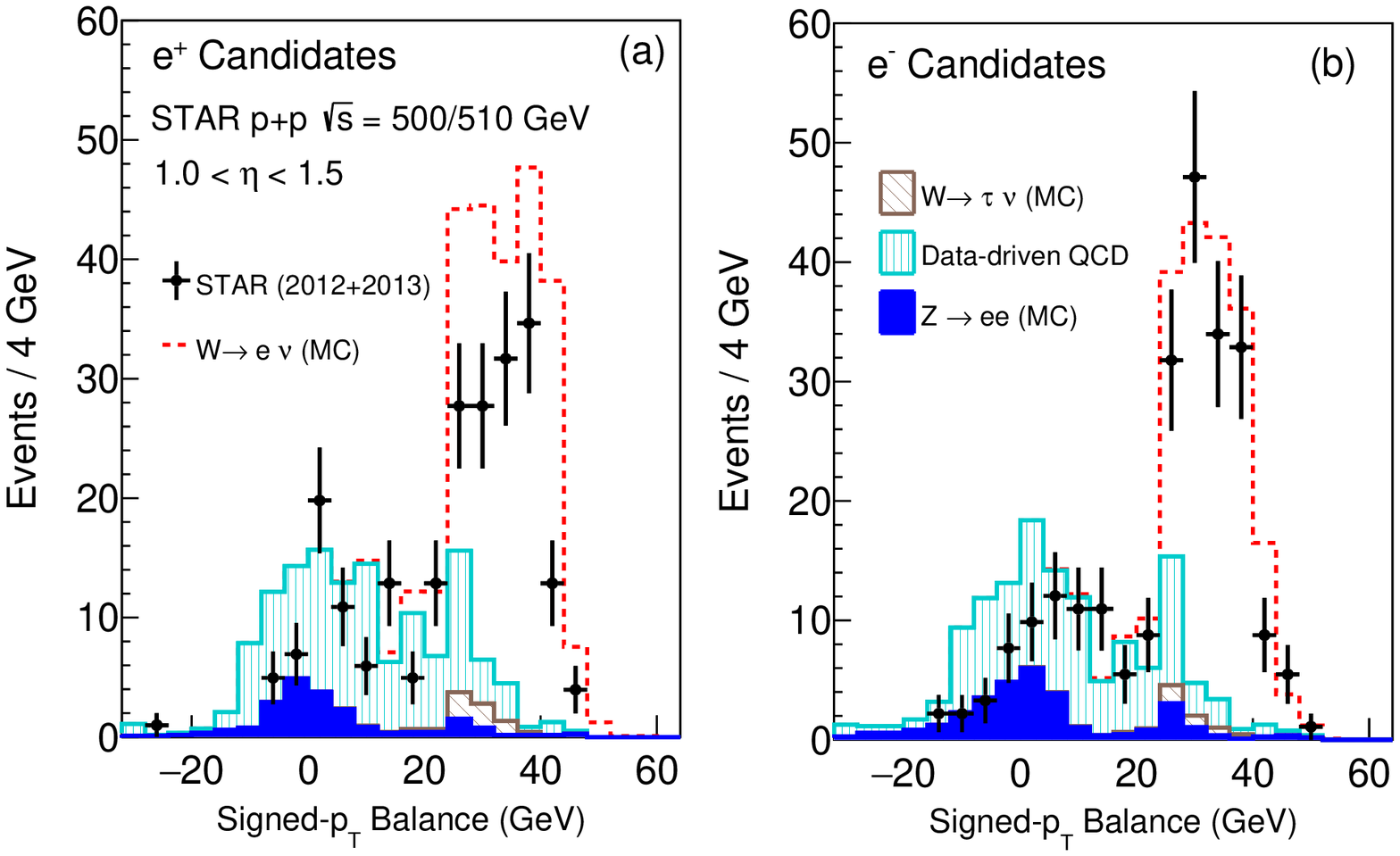}
\caption{\label{fig:EEMCbkgd}Signal and background signed-$p_T$ balance distributions for positron (a) and electron (b) candidates in the EEMC. The background contributions are shown as stacked histograms, where the solid blue and brown diagonal histograms correspond to the electroweak residual backgrounds from $Z\rightarrow ee$ and $W\rightarrow \tau\nu$ decay channels, respectively. The vertical cyan histograms correspond to the residual QCD contributions estimated from the data driven method. The red dashed histogram shows the $W\rightarrow e\nu$ signal along with all estimated background contributions and is compared to the data, the black markers. The vertical error bar on the data represents the statistical uncertainty and the horizontal bar shows the bin width.}
\end{figure}
The charge-sign associated with the lepton candidates is determined based on the curvature of their tracks measured in the TPC and STAR's magnetic field. The yield for a particular charge-sign in the BEMC is determined by fitting the $Q_e \cdot E^e_T/p_T$ distribution between $\pm 3.0$, where $Q_e$ is the charge-sign of the $e^\pm$ candidate determined from the curvature of its reconstructed track. Figure~\ref{fig:BEMCbkgd} shows the $E^e_T$ distributions for the $e^\pm$ decay candidates from the studied $W^\pm$ bosons decay channels, measured in the BEMC. The Jacobian peak in these distributions can clearly be seen between $30$ GeV and $40$ GeV. The electron and positron yields in the EEMC are also determined by fitting the $Q_e \cdot E^e_T/p_T$ distribution. Figure~\ref{fig:EEMCbkgd} shows the signed-$p_T$ balance distribution for $e^+$(left panel) and $e^-$(right panel) $W^\pm$ decay candidates in the EEMC.  Final $W$ candidates in the BEMC and EEMC are required to fall within the range $25$ GeV $< E^e_T < 50$ GeV. The details of the fits used to extract the $e^\pm$ yields and background estimates for these distributions will be discussed in Sec.~\ref{sec:background}. 

\subsection{$Z$ Candidate Event Selection}\label{sec:zcandidate}   
$Z\rightarrow e^+e^-$ candidate events can be selected by finding isolated $e^+e^-$ pairs. The isolated $e^\pm$ candidates were found using the isolation criteria discussed in Sec.~\ref{sec:isolationcuts}, with a slight modification to some of the isolation requirement values. For the $e^\pm$ candidates the ratio $E^e_T$ to the energy in the surrounding $4 \times 4$ cluster was required to be $95$\% and $E^e_T/E^{\Delta R>0.7}_T$ was required to be greater than $88$\%. 
\begin{figure}[tbh]
\center
\includegraphics[width = 1\columnwidth,keepaspectratio]{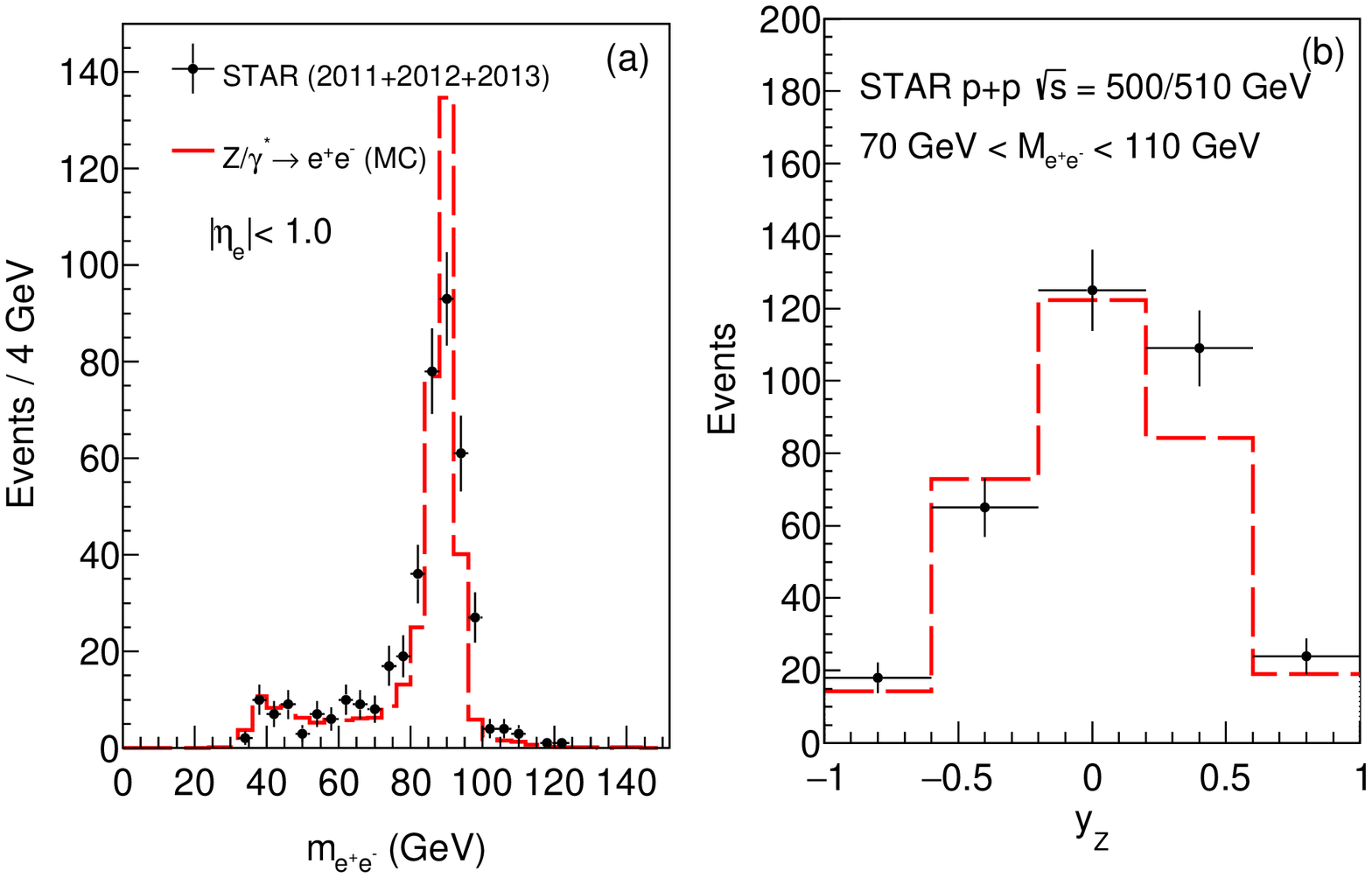}
\caption{\label{fig:ZRecon}Panel (a) shows the distribution of the reconstructed invariant mass from $Z$ decay candidates compared to $Z/\gamma^* \rightarrow e^+e^-$ MC distribution. Panel (b) shows the number of $Z$ candidate events plotted against the reconstructed rapidity and compared to the MC distribution. The red dashed histogram shows the $Z\rightarrow e^+e^-$ MC signal and is compared to the data, the black markers. The vertical error bar on the data represents the statistical uncertainty and the horizontal bar shows the bin width. The asymmetry in the MC between negative and positive $y_Z$ in (b) can be attributed to the rapidity asymmetry in the efficiencies, seen in Fig.~\ref{fig:Eff} (d), since these events have not yet been corrected for detector and cut efficiencies.}
\end{figure}
In addition to the isolation cuts, $Z$ decay $e^\pm$ candidates were also required to have a $p_T > 15$ GeV, $|\eta_e|<1.0$, and a charge-weighted $E^e_T/p_T$  satisfying $|Q_e\cdot E^e_T/p_T| \le 3.0$. Finally, by reconstructing the invariant mass of the $e^+e^-$ pairs, a fiducial cut was placed around the $Z$ mass covering the range $70$ GeV $\le m_{e^+e^-} \le 110$ GeV. The reconstructed invariant mass distribution is shown in Fig.~\ref{fig:ZRecon} (a), where the $Z/\gamma^* \rightarrow e^+e^-$ MC distribution is also shown for comparison. One can clearly see the $Z$ signal peak around the mass of the $Z$ near $91$ GeV. Figure~\ref{fig:ZRecon} (b) shows the number of $Z$ candidates plotted against the reconstructed $Z$-boson rapidity. Good agreement is found between the data and MC distributions.
\section{Signal and Background Estimates}
\label{sec:background}
\subsection{$W$ Signal and Background Estimation}
\begin{figure}[h]
\center
\includegraphics[width = 1\columnwidth,keepaspectratio]{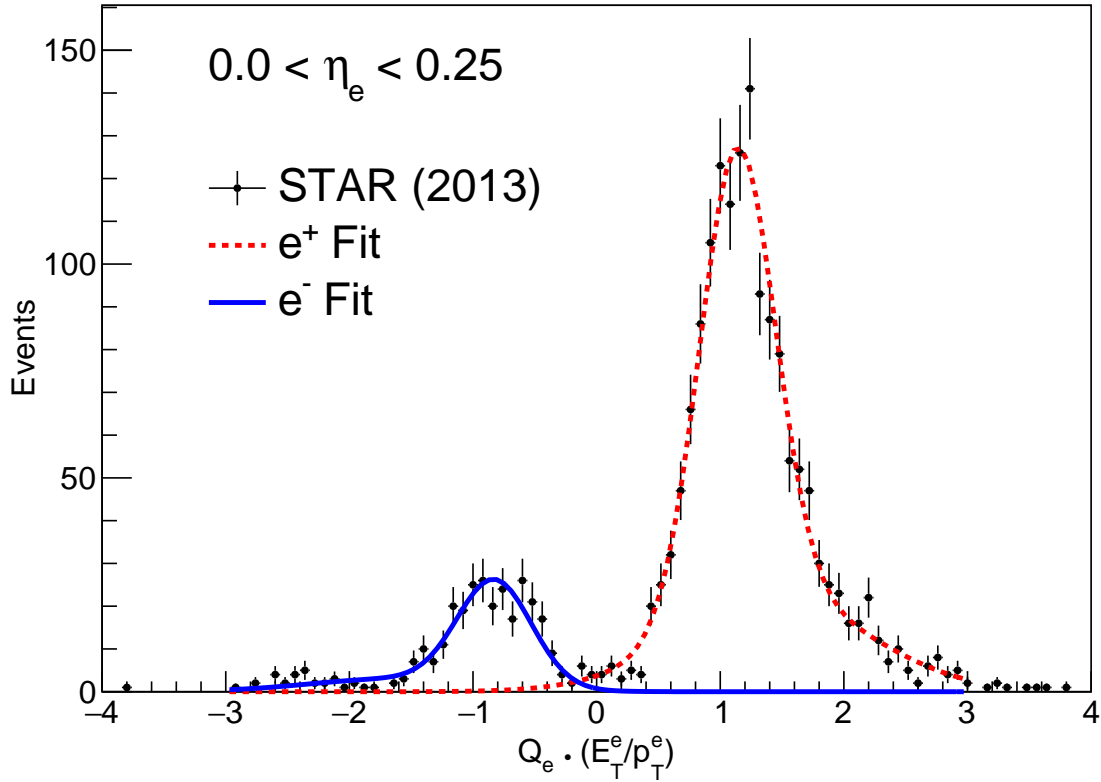}
\caption{\label{fig:EoverP_BEMC}A four-Gaussian function fit to measured $0.0 < \eta_e < 0.25$ BEMC $Q_e\cdot E^e_T/p_T$ distribution using the log-likelihood method. The colored lines show individual $e^+$(red dashed) and $e^-$(solid blue) double-Gaussian distributions resulting from the four-Gaussian fit.}
\end{figure}

The $e^\pm$ yields were determined by fitting the charge-weighted $E^e_T/p_T$ distribution. The fits were done for each of the eight pseudorapidity bins, separately for each of the three data sets. Following the fit procedures used in Ref.~\cite{Adam:2018bam}, the distributions were fitted using two double-Gaussian template shapes, determined from MC. To adequately describe the data, one Gaussian function was used to determine the $E^e_T/p_T$ distribution from the $W \rightarrow e\nu$ signal, while the other Gaussian function was used to describe the tails. The former resulted in a narrower distribution than the latter. The amplitudes were fitted to the data, using the log-likelihood method, along with the width and peak position of the narrower Gaussian in each of the templates. The remaining parameters were fixed based on the MC fit. Figure~\ref{fig:EoverP_BEMC} shows the fit result for the $0.0 \le \eta_e \le 0.25$ pseudorapidity bin from the 2013 data set. The red dashed line represents the fit to the positron distribution, while the blue solid line shows the fit to the electron distribution. This fit result is representative of the fits performed in the other pseudorapidity bins and other data sets. The positron and electron yields were determined by integrating the respective double-Gaussian function derived from the four-Gaussian function total fit. 

Two main sources which lead to misidentified $W$ candidates in $W\rightarrow e\nu$ decays are from electroweak and partonic processes~\cite{STAR:2011aa,Aggarwal:2010vc,Adamczyk:2014xyw,Adamczyk:2015gyk,Adam:2018bam}. A combination of MC samples and data was used to estimate these backgrounds. The background estimation procedure we used follows the same procedure detailed in Ref.~\cite{Adam:2018bam}. We then applied the estimated background fractions to the yields found from the fits discussed above. 

%BEMC
%--Z Bkgd
Two sources of electroweak backgrounds in $W$ decay are from $W\rightarrow \tau\nu$ and $Z\rightarrow e^+e^-$, where one of the $Z$ decay particles goes undetected due to either detector inefficiencies or acceptance effects. The contribution of these processes to the $W\rightarrow e\nu$ yield was estimated using MC samples described in Sec.~\ref{sec:data}. 
%A systematic uncertainty was assigned to this estimate based on the $9$\% uncertainty of the luminosity measurement that went into normalizing the simulated decay distributions. 

%--QCD Bkgd
The residual QCD dijet background is mainly due to one of the jets pointing to a region outside of the STAR acceptance. For the mid-pseudorapidity region (BEMC) this background had two contributions~\cite{STAR:2011aa,Adamczyk:2014xyw,Adam:2018bam}. The first contribution, referred to as the~``second EEMC" background, uses the instrumented EEMC in the pseudorapidity region $1.1 < \eta < 2$ to estimate the background associated with $e^\pm$ candidates that have an opposite-side jet fragment outside the detector region $-2 < \eta <-1.1$. The second contribution, referred to as the~``data-driven QCD" background, estimates the QCD background where one of the dijet fragments escapes through the uninstrumented regions at $|\eta| > 2$. This procedure looks at events that pass all $W$ selection criteria, but fail the signed-$p_T$ balance requirement. The background distribution was determined by normalizing the $E_T$ distribution to the $W$ candidate $E^e_T$ distribution between 16 GeV and 20 GeV after all other background contributions and the $W$ MC signal were removed. Both of these procedures are detailed in Ref.~\cite{STAR:2011aa}. Figure~\ref{fig:BEMCbkgd} shows the measured $W^+$ and $W^-$ yields as a function of $E^e_T$ over the integrated BEMC pseudorapidity range ($|\eta_e|<1$) along with the various estimated background contributions and the MC signal distribution for the combined 2011, 2012, and 2013 data sets. The systematic uncertainty associated with the data-driven QCD method was estimated by varying the signed-$p_T$ balance cut value and the $E_T$ window used to normalize the QCD background. The signed-$p_T$ balance cut was varied between $5$ GeV and $25$ GeV, while the $E_T$ normalization window was varied between $16$ GeV and $23.5$ GeV. Events which fail the signed-$p_T$ balance cut, which are dominated by dijet events, are used to estimate the QCD background where dijets escape detection at $|\eta|>2$. However, dijet events selected using this method, contain jets that were detected in the region $-1 < \eta < 2$. To account for the difference in the dijet cross sections, a PYTHIA study looking at hard partonic processes was carried out comparing the dijet cross section distributions in the regions $|\eta|<1$ and $|\eta|>2$. The relative difference between the two, with respect to the mid-pseudorapidity distribution, $\sim 43$\%, was taken as an additional systematic uncertainty to the QCD background yield found using the data-driven QCD method. The average background contributions were found to be several percent of the total $W$ yields, and the background to signal ratio for each process is listed in Table~\ref{tbl:bkg_BEMC}. 

The EEMC measurements have a greater likelihood of having the charge-sign misidentified compared to the BEMC. Intermediate pseudorapidity tracks miss the outer radius of the TPC and thus tracking resolution is degraded resulting in broader charge-weighted $E^e_T/p_T$ distributions and larger charge contamination compared to distributions measured at mid-pseudorapidity. It was found that the data could be well described using a two-Gaussian function where each Gaussian function described the particular charge's $E^e_T/p_T$ distribution. As a result the charge separated yield was determined by fitting the EEMC $E^e_T/p_T$ distribution with a two-Gaussian function using the log-likelihood method and integrating over the resulting single Gaussian functions for each $e^\pm$ yield. The results of this fit are shown in Fig.~\ref{fig:EoverP_EEMC}.~The electron and positron contributions resulting from the two-Gaussian total fit are shown as the blue solid and red dashed lines, respectively. A systematic uncertainty of about 3\% was estimated by varying the two-Gaussian fitting limits by $\pm0.3$.     
%--Z and QCD Bkgd
The estimation of background contributions in the EEMC followed a procedure similar to the one used for the BEMC. 
The determined background fractions were then applied to the yields determined from the $E^e_T/p_T$ fit. The dominant background sources again resulted from electroweak ($W\rightarrow \tau\nu$ and $Z\rightarrow e^+e^-$) and the hard partonic processes. The residual electroweak decay contamination was determined from MC samples, while the QCD background was estimated using only the data-driven QCD method. The residual QCD backgrounds were estimated using the ESMD, where the isolation parameter $R_{ESMD}$ was required to be less than $0.6$ for QCD background candidates. This sample was then normalized to the measured $W$ candidate signed-$p_T$ balance distribution between $-8$ GeV and $14$ GeV, where the QCD background dominates. Figure~\ref{fig:EEMCbkgd} shows the measured $W^+$ and $W^-$ yields as a function of signed-$p_T$ balance, along with the estimated backgrounds and MC signal distribution for the combined 2012 and 2013 data sets. The data-driven QCD systematic uncertainty was determined by varying the $R_{ESMD}$ cut value between $0.4$ and $0.55$. Furthermore the signed-$p_T$ balance window, which was used to normalize the QCD background, was varied between $-4.0$ GeV and $22.0$ GeV to assess the data-driven QCD's sensitivity to the normalization window. Table~\ref{tbl:bkg_EEMC} summarizes the various background estimates in the EEMC.   
\begin{center}
\begin{table}[h]
%%\begin{longtable}[h]
\caption{\label{tbl:bkg_BEMC} Combined 2011, 2012, and 2013 background to signal ratio for $W^+$ and $W^-$ between $25$ GeV $ < E^e_T < 50$ GeV and $|\eta_{e}|<1$.}
\begin{ruledtabular}
\begin{tabular}{ccccc}
Background & $W\rightarrow \tau\nu$ (\%) & $Z\rightarrow e^+e^-$ (\%) & Data-driven QCD (\%)& Second EEMC QCD(\%)\\
\hline
B/S ($W^+$) & $2.1 \pm 0.1$ (stat.) & $1.1 \pm 0.1$ (stat.) & $2.1 \pm 0.1$ (stat.) $\pm 1.2$ (sys.) & $4.2  \pm 0.2$ (stat.)\\
B/S ($W^-$) & $2.1 \pm 0.2$ (stat.) & $3.8 \pm 0.4$ (stat.) & $4.6 \pm 0.3$ (stat.) $\pm 2.4$ (sys.) & $10.9 \pm 0.6$ (stat.)\\
\end{tabular}
\end{ruledtabular}
\end{table}

\begin{figure}[h]
\center
\includegraphics[width = 1\columnwidth,keepaspectratio]{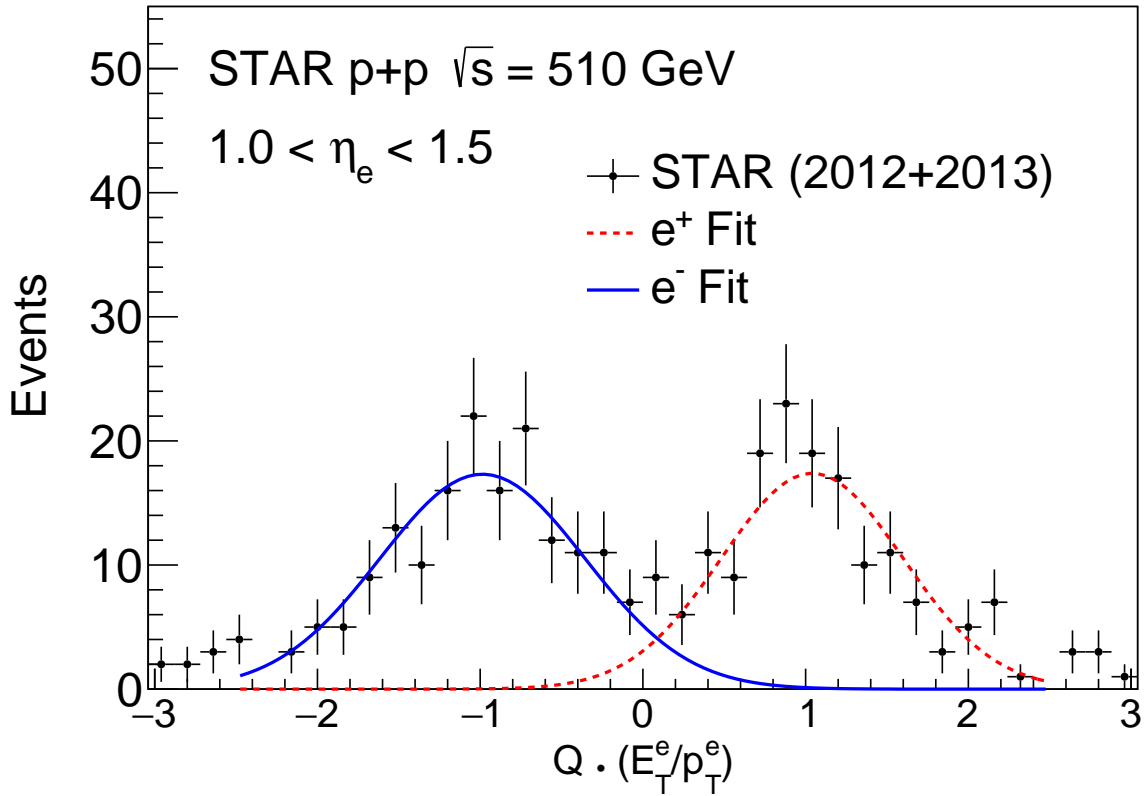}
\caption{\label{fig:EoverP_EEMC}Double Gaussian fit to measured EEMC $Q_e\cdot E^e_T/p_T$ distribution using the log-likelihood method. The colored lines show individual $e^+$(red dashed) and $e^-$(solid blue) Gaussian distributions resulting from the double-Gaussian fit.}
\end{figure}

\begin{table}[h]
\caption{\label{tbl:bkg_EEMC} Combined 2012 and 2013 background to signal ratio for $W^+$ and $W^-$ for $25$ GeV $ < E^e_T < 50$ GeV, $R_{ESMD} > 0.7$, and signed-$p_T$ balance $> 20$ GeV in $1.0 < \eta_{e} < 1.5$. Not shown in the table is the 3\% uncertainty associated with the fit to the charge-weighted $W$ yields.}
\begin{ruledtabular}
\begin{tabular}{cccc}
Background & $W\rightarrow \tau\nu$ (\%) & $Z\rightarrow e^+e^-$ (\%) & Data-driven QCD (\%)\\
\hline
B/S ($W^+$) & $3.9 \pm 0.5$ (stat.) & $2.3 \pm 0.4$ (stat.) & $11.3\pm 2.6$ (stat.) $\pm 2.0$ (sys.) \\
B/S ($W^-$) & $2.1 \pm 0.3$ (stat.) & $3.7 \pm 0.5$ (stat.) & $7.7 \pm 1.8$ (stat.) $\pm 1.5$ (sys.) \\ 
\end{tabular}
\end{ruledtabular}
\end{table}
\end{center}

\subsection{$Z$ Signal and Background Estimation}
Due to the requirement of having a pair of oppositely charged, high-$E_T$, and isolated $e^+$ and $e^-$, the background in $Z\rightarrow e^+e^-$ is expected to be small. The background was estimated by comparing the number of lepton pairs with the same-charge sign, which passed all $Z$ candidate selection criteria, to those which had opposite-charge sign. This background was found to be just under 4\% in our combined data sets. Background corrections were applied to each rapidity bin for each of the three data sets by subtracting the number of same-charge sign events which passed the $Z$ candidate criteria from the number of opposite-charge sign $Z$ candidates. 
\section{Efficiencies}
\label{sec:eff}
The measured fiducial cross sections can be written as 
\begin{equation}\label{eq:fid}
\sigma^{fid}_{W} = \frac{N^{obs}_{W}-N^{bkgd}_{W}}{\mathcal{L}\cdot \varepsilon_{W}},
\end{equation}
\noindent where $N^{obs}_{W}$ is the number of observed $W$ candidates within the defined kinematic acceptance that meet the selection criteria specified in Sec.~\ref{sec:recon}. $N^{bkgd}_{W}$ is the total number of background events within the defined kinematic acceptance, as described in Sec~\ref{sec:background}. $\mathcal{L}$ is the total integrated luminosity, and $\varepsilon_{W}$ is the efficiency that needs to be applied to correct for detector and cut effects. Equation~\ref{eq:fid} also describes the $Z$ fiducial cross section, $\sigma^{fid}_{Z}$, with the replacement of $W$ related quantities with the $Z$ related quantities. 

The $W$ and $Z$ efficiencies were computed in the same manner as in Ref.~\cite{STAR:2011aa}. The efficiencies were defined as the ratios between the number of $W$($Z$) boson decay candidates satisfying selection criteria to all those $W$($Z$) bosons falling within the STAR fiducial acceptance.

\begin{center}
\begin{figure}[h]
\includegraphics[width = 1.0\columnwidth,keepaspectratio]{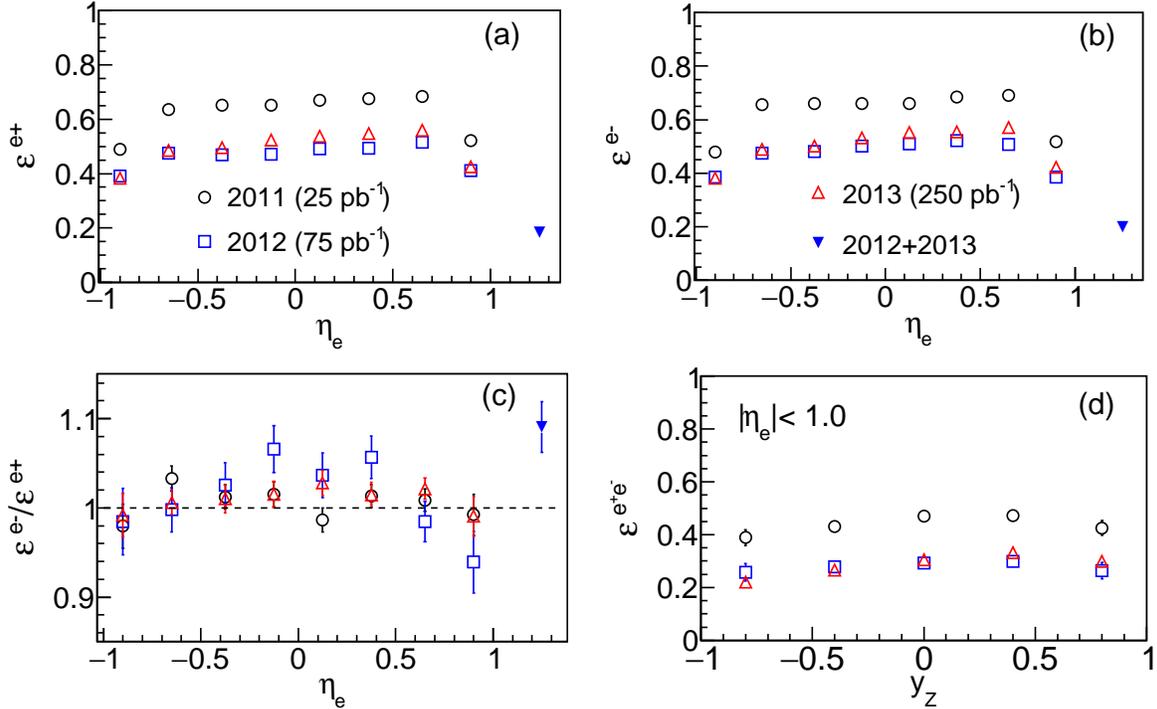}
\caption{\label{fig:Eff}Individual data set efficiencies for: positron (a) and electron (b) $W^\pm$ decay candidates plotted as a function of pseudorapidity. Panel (c) shows the $e^-/e^+$ efficiency ratio vs. pseudorapidity. Panel (d) shows the efficiency for $Z$ decay candidates vs. rapidity.}
\end{figure}
\end{center}

The $W$ candidate efficiencies for each of the three data sets are plotted in Fig.~\ref{fig:Eff} (a) for positron and (b) electron candidates as a function of pseudorapidity. Comparing the $W$ efficiencies between the three data sets, one can clearly see a larger efficiency for the 2011 data set. This is due primarily to a lower instantaneous luminosity relative to the 2012 and 2013 data sets. The higher instantaneous luminosity leads to larger pile-up in the TPC, resulting in less efficient track reconstruction. The 2013 data set used a new track reconstruction algorithm which resulted in a more efficient track reconstruction. This counteracted much of the efficiency loss that would come with increasing the instantaneous luminosity, allowing for efficiencies that are comparable to those found in the 2012 data set. The positron and electron efficiencies amongst each data set are comparable as can be seen in Fig.~\ref{fig:Eff} (c), which plots the ratio $\varepsilon_{W^-}/\varepsilon_{W^+}$ as a function of pseudorapidity. The relatively small offset from one shows that the efficiency corrections will have a small effect to the $\sigma^{fid}_{W^+} / \sigma^{fid}_{W^-}$ measurement. Figure~\ref{fig:Eff} (d) shows the $Z$ efficiencies computed for the three data sets as a function of rapidity. The $Z$ efficiencies are overall lower than the $W$ efficiencies, since for $Z$ candidates we required two reconstructed tracks.

There were two sources of systematic uncertainties associated with the efficiencies, the estimation of which was based on a previous STAR analysis~\cite{STAR:2011aa}. The first is associated with TPC track reconstruction efficiency for $W$ or $Z$ candidates. Based on past analyses,  the uncertainty of $5$\% and $10$\% was used for the $W$ and $Z$ tracking efficiency, respectively. The second systematic uncertainty is related to how well the BEMC and EEMC energy scales are known. This systematic uncertainty was propagated to the efficiencies by varying the BEMC and EEMC energy scale by its gain uncertainty of $5$\%. However, when evaluating the cross-section ratios (Sec.~\ref{sec:ratios}) many of these systematic uncertainties either partially or completely cancel.

\section{$W$ and $Z$ Cross Sections}
\label{sec:xsec}
\subsection{$W$ and $Z$ Differential Cross Sections}\label{sec:diffxsec}
\begin{center} 
\begin{figure}[!h]
\includegraphics[width = 0.8\columnwidth,keepaspectratio]{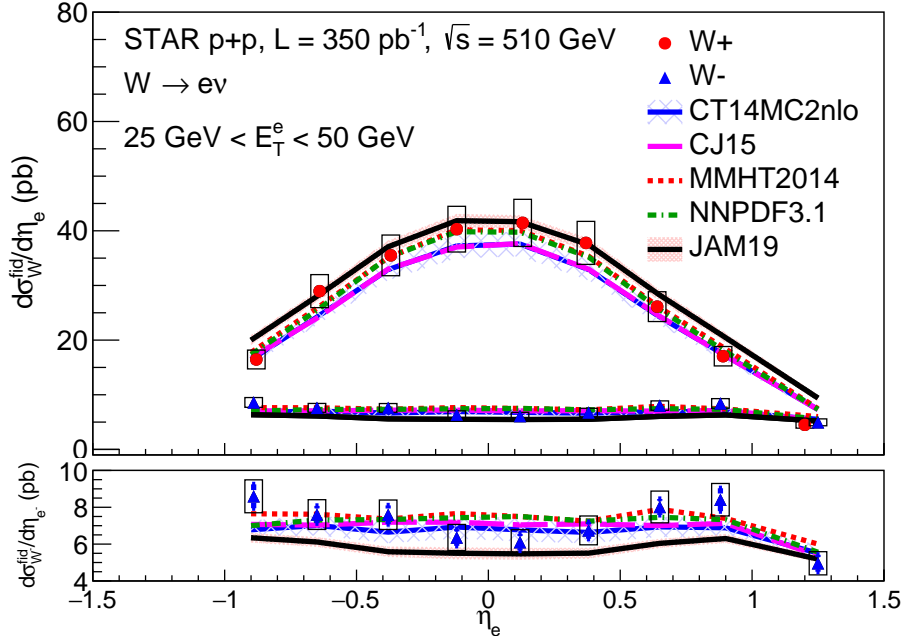}
\caption{\label{fig:dW}The measured $d\sigma^{fid}_{W^+}/d\eta_{e^+}$ (closed circle markers) and $d\sigma^{fid}_{W^-}/d\eta_{e^-}$ (closed triangle markers) for the combined data sets (2011-2013) are plotted as a function of $\eta_{e}$. The bottom panel shows $d\sigma^{fid}_{W^-}/d\eta_{e^-}$ when zooming in on the vertical axis. FEWZ~\cite{PhysRevD.86.094034} was used to compare various NLO PDF sets (CT14MC2nlo~\cite{Hou2017}, CJ15~\cite{Accardi:2016qay}, MMHT14~\cite{Harland-Lang2015}, NNPDF3.1~\cite{Ball2017}, and JAM19~\cite{Sato:2019yez}) to the measured differential cross sections.}
\end{figure}
\end{center}

\begin{figure}[!h]
\center
\includegraphics[width = 0.8\columnwidth,keepaspectratio]{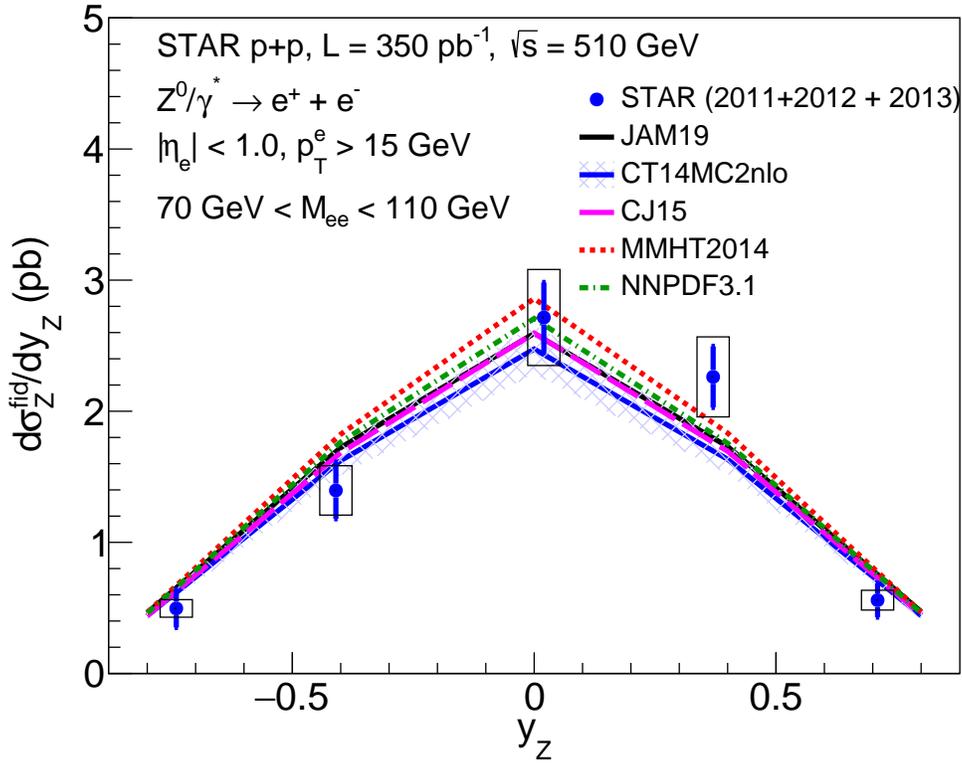}
\caption{\label{fig:dZ}The measured $d\sigma^{fid}_{Z}/dy_Z$ for the combined data sets (2011-2013) is plotted against the $Z$ rapidity, and compared to theory calculations done using FEWZ~\cite{PhysRevD.86.094034} for several different NLO PDF sets (CT14MC2nlo~\cite{Hou2017}, CJ15~\cite{Accardi:2016qay}, MMHT14~\cite{Harland-Lang2015}, NNPDF3.1~\cite{Ball2017}, and JAM19~\cite{Sato:2019yez}).}
\end{figure}

Using the selected $W$ and $Z$ candidates discussed in Sec.~\ref{sec:recon}, correcting them for background contamination following Sec.~\ref{sec:background}, and finally applying the efficiency corrections computed in Sec.~\ref{sec:eff}, Eq.~\ref{eq:fid} can be used to compute the differential cross sections $d\sigma^{fid}_{W^\pm}/d\eta_{e^\pm}$ and $d\sigma^{fid}_{Z}/dy_Z$. 
The measured differential cross sections $d\sigma^{fid}_{W^+}/d\eta_{e^+}$ and $d\sigma^{fid}_{W^-}/d\eta_{e^-}$ were obtained in nine pseudorapidity bins, that cover the range $-1.0 < \eta_e < 1.5$. Figure~\ref{fig:dW} shows the results for the combined data sets, where the statistical uncertainty is given by the error bars and the total systematic uncertainties are represented by the boxes surrounding the respective data points. These boxes do not represent a horizontal uncertainty. The bottom panel of Fig.~\ref{fig:dW} modifies the range of the vertical scale to see better the trend of the $W^-$ differential cross section. Using FEWZ~\cite{PhysRevD.86.094034} in combination with LHAPDF~\cite{Buckley2015}, the differential cross sections were evaluated using several PDF sets: CT14MC2nlo~\cite{Hou2017}, CJ15~\cite{Accardi:2016qay}, MMHT2014~\cite{Harland-Lang2015}, NNPDF 3.1~\cite{Ball2017}, and JAM19~\cite{Sato:2019yez}. The CT14MC2nlo PDF set contains 1000 replicas and the uncertainty used in the PDF band represents the RMS value in the quantity evaluated from the 1000 replicas. The JAM19 PDF set typically yields smaller values for $W^-$ compared to our measurements. This will result in larger $W^+/W^-$ cross-section ratios compared to our measured values. Table~\ref{tbl:dW} lists the $W^\pm$ differential cross sections and their associated uncertainties that are shown in Fig.~\ref{fig:dW}. Figure~\ref{fig:dZ} shows the combined 2011, 2012, and 2013 measured $Z$ differential cross section, $d\sigma^{fid}_{Z}/dy_Z$, as a function of the rapidity. The $Z$ differential cross section was binned in five equally spaced $Z$ rapidity bins. The statistical uncertainties are represented by the error bars, while the total systematic uncertainties are displayed as boxes around the data points. These boxes represent only a vertical uncertainty. The experimental results are compared to theory calculations done using FEWZ~\cite{PhysRevD.86.094034} for several different PDF sets (CT14MC2nlo~\cite{Hou2017}, CJ15~\cite{Accardi:2016qay}, MMHT14~\cite{Harland-Lang2015}, NNPDF3.1~\cite{Ball2017}, and JAM19~\cite{Sato:2019yez}). The cross section values, shown in Fig.~\ref{fig:dZ}, are provided in Table~\ref{tbl:dZ}.

\begin{table}[!h]
\caption{\label{tbl:dW} Combined (2011,2012, and 2013) results for differential cross sections, $d\sigma^{fid}_{W^\pm}/d\eta_{e}$, binned in $e^\pm$ pseudorapidity bins, requiring that $-1 < \eta_e < 1.5$ and $25$ GeV $< E^e_{T} < 50$ GeV. The columns labeled ``Stat.'' and ``Eff.'' represent the statistical uncertainty and the systematic uncertainty estimated from the efficiencies, respectively. The later is dominated by the 5\% uncertainty in the tracking efficiency, which is common to all the measurements. The column ``Sys.'' includes all remaining systematic uncertainties, with the exception of the luminosity. The 9\% uncertainty associated with the luminosity measurement is not included in the table.}
\begin{ruledtabular}
\begin{tabular}{cccccc}
$\eta_{e}$ Range & $<\eta_{e^+}>$ & $d\sigma^{fid}_{W^+}/d\eta_{e^+}$ (pb) & Stat. (pb) & Sys. (pb) & Eff. (pb)\\
\hline
 $-1.00$, $-0.80$ & $-0.88$ & $16.5$ & $0.9$ & $0.3$ &$0.8$ \\
 $-0.80$, $-0.50$ & $-0.64$ & $29.0$ & $0.8$ & $0.4$ &$1.5$ \\
 $-0.50$, $-0.25$ & $-0.37$ & $35.5$ & $1.0$ & $0.6$ &$1.8$ \\
 $-0.25$, $0.00$ & $-0.12$ & $40.3$ & $1.0$ & $0.3$ &$2.1$ \\
 $0.00$, $0.25$  & $0.13$ & $41.4$ & $1.0$ & $0.4$ &$2.1$ \\
 $0.25$, $0.50$  & $0.37$ & $37.8$ & $1.0$ & $0.4$ &$1.9$ \\
 $0.50$, $0.80$  & $0.64$ & $26.1$ & $0.7$ & $0.4$ &$1.3$ \\
 $0.80$, $1.00$  & $0.89$ & $17.1$ & $0.9$ & $0.2$ &$0.9$ \\
 $1.00$, $1.50$  & $1.20$ & $4.5$ & $0.5$ & $0.2$ &$0.4$  \\
%\hline
\hline
$\eta_{e}$ Range & $<\eta_{e^-}>$ & $d\sigma^{fid}_{W^-}/d\eta_{e^-}$ (pb) & Stat. (pb) & Sys. (pb) & Eff. (pb)\\
\hline
 $-1.00$, $-0.80$ & $-0.89$ & $8.6$ & $0.6$ & $0.1$ &$0.4$ \\
 $-0.80$, $-0.50$ & $-0.65$ & $7.6$ & $0.5$ & $0.2$ &$0.4$ \\
 $-0.50$, $-0.25$ & $-0.38$ & $7.6$ & $0.5$ & $0.2$ &$0.4$ \\
 $-0.25$, $0.00$ &$-0.12$ & $6.4$ & $0.5$ & $0.3$ &$0.3$ \\
 $0.00$, $0.25$  &$0.12$ & $6.1$ & $0.5$ & $0.3$ &$0.3$  \\
 $0.25$, $0.50$  & $0.38$ & $6.7$ & $0.5$ & $0.4$ &$0.3$ \\
 $0.50$, $0.80$  & $0.65$ & $8.0$ & $0.4$ & $0.2$ &$0.4$  \\
 $0.80$, $1.00$  &$0.88$ & $8.4$ & $0.6$ & $0.1$ &$0.4$  \\
 $1.00$, $1.50$  &$1.25$ & $5.0$ & $0.5$ & $0.2$ &$0.4$  \\
\end{tabular}
\end{ruledtabular}
\end{table}

%\noindent Figure~\ref{fig:dZ} shows the combined 2011, 2012, and 2013 measured $Z$ differential cross section, $d\sigma^{fid}_{Z}/dy_Z$, as a function of the rapidity. The $Z$ differential cross section was binned in five equally spaced $Z$ rapidity bins. The statistical uncertainties are represented by the error bars, while the total systematic uncertainties are displayed as boxes around the data points. These boxes represent only a vertical uncertainty. The experimental results are compared to theory calculations done using FEWZ~\cite{PhysRevD.86.094034} for several different PDF sets (CT14MC2nlo~\cite{Hou2017}, CJ15~\cite{Accardi:2016qay}, MMHT14~\cite{Harland-Lang2015}, NNPDF3.1~\cite{Ball2017}, and JAM19~\cite{Sato:2019yez}). The cross section values, shown in Fig.~\ref{fig:dZ}, are provided in Table~\ref{tbl:dZ}. 

\begin{table}[tbh]
\caption{\label{tbl:dZ} Combined (2011,2012, and 2013) results for the differential cross section, $d\sigma^{fid}_{Z}/dy_Z$, binned in rapidity bins, requiring that $|\eta_e|<1$, $|y_Z| < 1$, $p^e_T > 15$ GeV, and $ 70$ GeV $< M_Z < 110$ GeV. The columns labeled ``Stat.'' and ``Eff.'' represent the statistical uncertainty and the systematic uncertainty estimated from the efficiencies, respectively. The later is dominated by the 10\% uncertainty in the tracking efficiency, which is common to all the measurements. The 9\% uncertainty associated with the luminosity measurement is not included in the table.}
\begin{ruledtabular}
\begin{tabular}{cccc}
$<y_Z>$ & $d\sigma^{fid}_{Z}/dy_Z$ (pb) & Stat. (pb) & Eff. (pb) \\
\hline
 $-0.74$ & $0.5$ & $0.1$ & $0.05$\\
 $-0.41$ & $1.4$ & $0.2$ & $0.14$\\
 $0.02$  & $2.7$ & $0.3$ & $0.27$\\
 $0.37$  & $2.3$ & $0.2$ & $0.23$\\
 $0.71$  & $0.6$ & $0.1$ & $0.06$\\
\end{tabular}
\end{ruledtabular}
\end{table}

\subsection{$W$ and $Z$ Total Cross Sections}
%Total Cross Section
\begin{table}[h]
\caption{\label{tbl:fid} Total fiducial cross section results for combined 2011, 2012, and 2013 data sets and their corresponding uncertainties. The columns labeled ``Stat.'' and ``Eff.'' represent the statistical uncertainty and the systematic uncertainty estimated from the efficiencies, respectively. The column ``Sys.'' includes all remaining systematic uncertainties, with the exception of the luminosity. The 9\% uncertainty associated with the luminosity measurement is not included in the table.}
\begin{ruledtabular}
\begin{tabular}{ccccc}
 & Value(pb) & Stat.(pb) & Sys.(pb) & Eff.(pb)\\
\hline
$\sigma^{fid}_{W^+}$ &$64.3$ & $0.7$ & $0.9$ & $3.4$\\
$\sigma^{fid}_{W^-}$ &$17.3$ & $0.4$ & $0.5$ & $0.9$\\
$\sigma^{fid}_{Z}$   &$3.0$ & $0.2$ & $0.0$  & $0.3$ \\
\end{tabular}
\end{ruledtabular}
\end{table}
The total fiducial cross sections can be obtained by integrating the differential cross sections. Table~\ref{tbl:fid} lists the values for the measured fiducial cross sections: $\sigma^{fid}_{W^+}$, $\sigma^{fid}_{W^-}$, and $\sigma^{fid}_{Z}$. From these, the total cross sections $\sigma^{tot}_{W^\pm} \cdot B\left(W\rightarrow e\nu\right)$ and $\sigma^{tot}_{Z/\gamma^*} \cdot B\left(Z/\gamma^*\rightarrow e^+e^-\right)$ can be calculated according to the relations

\begin{equation}\label{eq:totxsecW}
\sigma^{tot}_{W^\pm}\cdot B(W\rightarrow e\nu) = \frac{\sigma^{fid}_{W^\pm}}{A_{W^\pm}}
\end{equation}

\begin{equation}\label{eq:totxsecZ}
\sigma^{tot}_{Z}\cdot B(Z\rightarrow e^+e^-) = \frac{\sigma^{fid}_{Z}}{A_{Z}},
\end{equation}  

\noindent where $A$ is a kinematic correction factor for the respective boson.
The kinematic correction factor, which is needed to account for the incomplete STAR kinematic acceptance, was determined for the $W^+$, $W^-$, and $Z$ bosons by using FEWZ in combination with LHAPDF and an assortment of PDF sets. FEWZ was used with the CT14MC2nlo~\cite{Hou2017} PDF, to compute fiducial $W^\pm$ and $Z$ cross sections, $(\sigma^{fid}_{W^\pm,Z})_{FEWZ}$, in a kinematic region that mimics the STAR detector. Cross sections were also computed using the full leptonic kinematic acceptance, $(\sigma^{tot}_{W^\pm,Z})_{FEWZ}$. The kinematic correction factor was then defined as 
\begin{equation}\label{eq:A}
B\cdot A_b = \left(\sigma^{fid}_{b}\right)_{FEWZ}/\left(\sigma^{tot}_{b}\right)_{FEWZ},
\end{equation}
where $b$ represents the respective boson, $W^\pm$ or $Z$, and $B$ is the corresponding the branching ratio, $W\rightarrow e\nu$ or $Z\rightarrow e^+e^-$. The kinematic correction factors calculated using the CT14MC2nlo PDF set are listed in Table~\ref{tbl:AccError}, along with their evaluated uncertainties. 

We considered two contributions to the kinematic correction factor uncertainty. The first contribution, $\delta A_{PDF}$, was on the CT14MC2nlo PDF set itself. To estimate this $A_{W^\pm}$ and $A_Z$ were computed for each replica. A Gaussian fit was made to each boson's kinematic correction factor distribution and the Gaussian width was taken as the uncertainty. The second contribution, $\delta A_{\alpha_s}$, assessed the effect of changing the $\alpha_s$ used in the PDF sets. This was estimated by computing the kinematic correction factor using the NNPDF3.1~\cite{Ball2017} PDF set with three different $\alpha_s$ values (0.116, 0.118, and 0.120). The average difference from $\alpha_s$ = 0.118 was used as an uncertainty. Table~\ref{tbl:AccError} summarizes the two uncertainty contributions and the final uncertainty associated with $A_{W^\pm,Z}$, which was propagated to the total cross section as a systematic uncertainty.    
\begin{table}[!h]
\caption{\label{tbl:AccError} Kinematic correction factors needed to compute the total cross sections and their uncertainties.}
\begin{ruledtabular}
\begin{tabular}{cccc}
Contrib. & $\delta A_{W^+}$ (\%)& $\delta A_{W^-}$ (\%)& $\delta A_{Z}$ (\%)\\
\hline
$\delta A_{PDF}$ & $0.9$ & $1.5$ & $1.6$ \\
$\delta A_{\alpha_s}$ & $0.8$ & $0.3$ & $0.6$ \\
\hline
Total Uncertainty & $1.2$ & $1.5$ & $1.7$ \\
\hline
& $A_{W^+}$ & $A_{W^-}$ & $A_{Z}$\\
\hline
& $0.45 \pm 0.01$ & $0.42 \pm 0.01$ & $0.35 \pm 0.01$ \\
\end{tabular}
\end{ruledtabular}
\end{table}

\begin{center}
\begin{figure}[!h]
\includegraphics[width = 1.0\columnwidth,keepaspectratio]{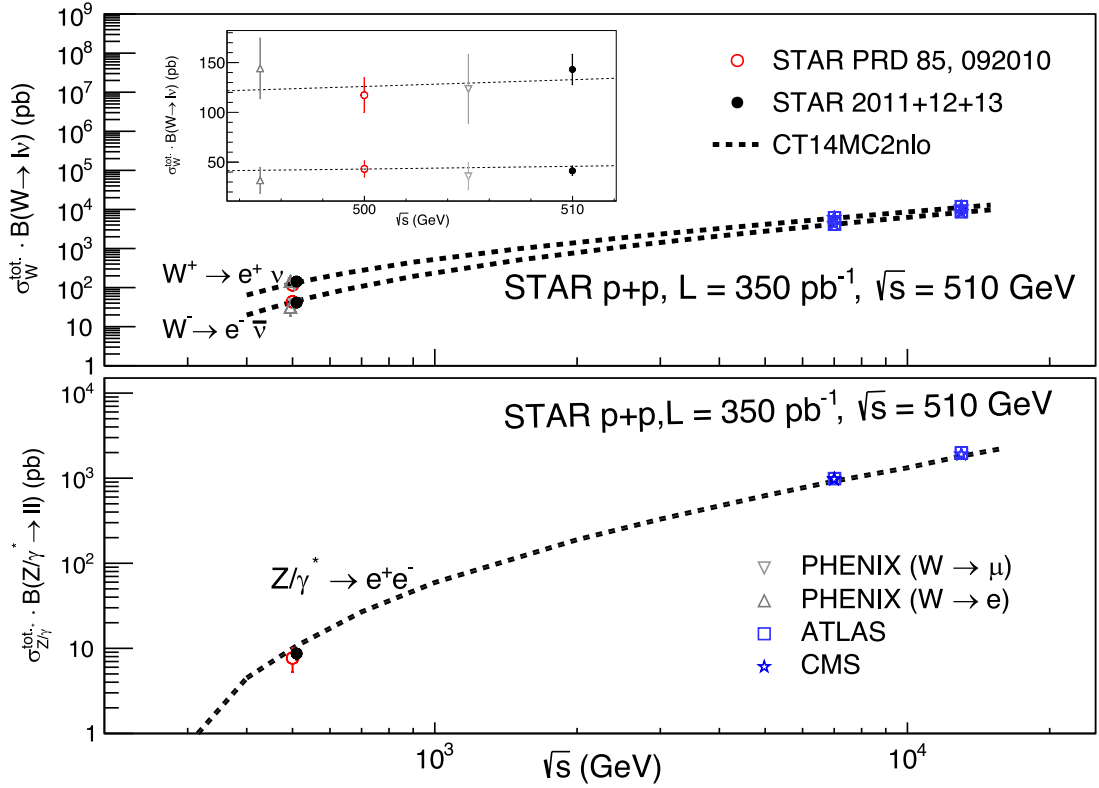}
\caption{\label{fig:TotalCrossSection}~The measured total $W^\pm$ and $Z$ cross sections for the combined STAR data sets (2011-2013). For clarity the PHENIX measurements are plotted at -5 GeV from $\sqrt{s}$ = 510 GeV ($W\rightarrow \mu$) and 500 GeV ($W\rightarrow e$), respectively. The inset plot in the upper panel highlights the STAR and PHENIX results ($\sqrt{s} \sim$ 500 GeV). For the $Z$ cross section, the STAR data uses a mass window of $70$ GeV $< m_{e^+e^-} < 110$ GeV, CT14MC2nlo and CMS use $60$ GeV $< m_{e^+e^-} < 120$ GeV, and ATLAS uses $66$ GeV $< m_{e^+e^-} < 116$ GeV. The dashed lines in the figure show the respective $W^\pm$ and $Z$ cross section curves computed using FEWZ and the CT14MC2nlo~\cite{Hou2017} PDF.}
\end{figure}
\end{center}

\begin{center}
\begin{figure}[!h]
\includegraphics[width = 1.0\columnwidth,keepaspectratio]{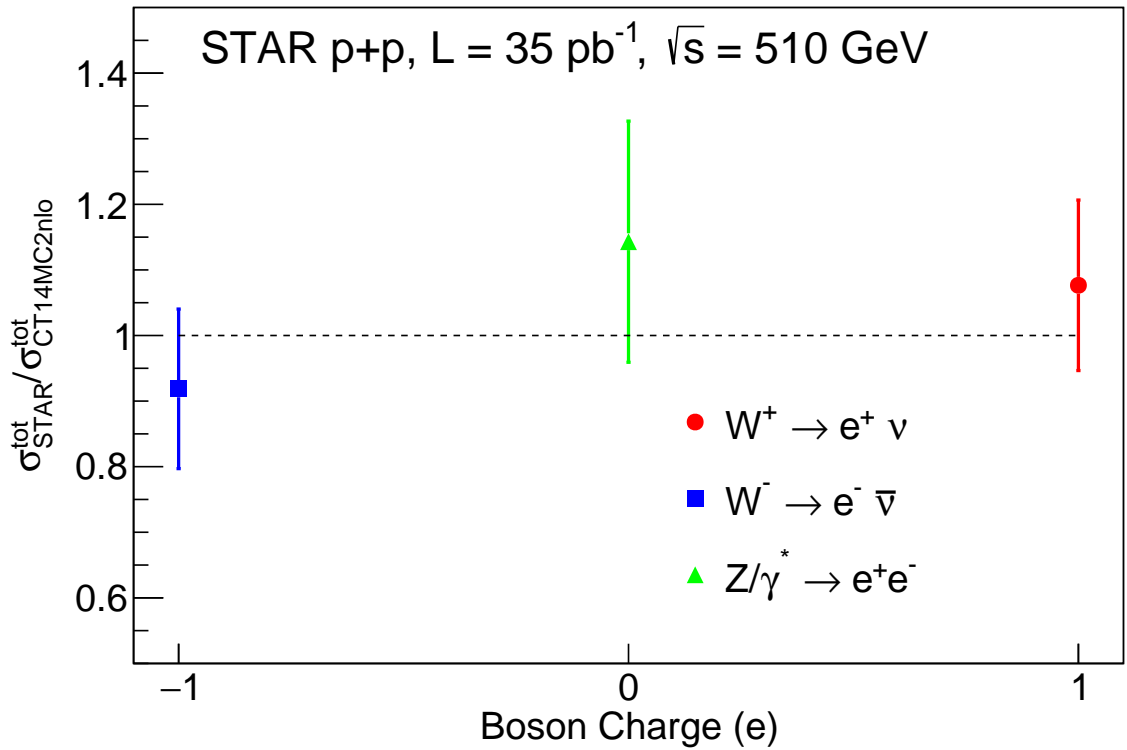}
\caption{\label{fig:TotalCrossSectionRatio} Ratio of the STAR calculated total cross sections to the total cross sections found using the CT14MC2nlo PDF set~\cite{PhysRevD.93.033006} versus the decay boson's charge. These comparisons place a mass window of $70$ GeV $< m_{e^+e^-} < 110$ GeV on the $Z$ cross section. The error bars shown here are the total uncertainties including contributions from the efficiency, luminosity, and PDF uncertainties.}
\end{figure}
\end{center}

The total $W^\pm$ and $Z$ cross sections were computed from the measured fiducial cross sections following Eqs.~\ref{eq:totxsecW} and~\ref{eq:totxsecZ}, and are shown in Fig.~\ref{fig:TotalCrossSection}. The top panel displays the $W^+$ and $W^-$ total cross sections, while the bottom panel shows the $Z$ total cross section.~Included for comparison are curves produced with FEWZ using the CT14MC2nlo~\cite{Hou2017} PDF set, as well as PHENIX~\cite{Adare:2010xa,Adare:2018csm} and previous STAR~\cite{STAR:2011aa} results at $\sqrt{s} =$ 500 and 510 GeV, and LHC data~\cite{Aad:2016naf,Aaboud:2016btc,Schott2014,CMS:2015ois} at larger $\sqrt{s} =$ 7 and 13 TeV. There is good agreement between this $W^\pm$ cross section measurement and those from previous STAR~\cite{STAR:2011aa} and PHENIX~\cite{Adare:2010xa,Adare:2018csm} analyses, which makes it difficult to distiguish them in the figure. As a result we have included in the figure a panel highlighting this region. Table~\ref{tbl:Xstot} lists the values of the combined 2011, 2012, and 2013 total cross sections and their associated uncertainties.~Figure~\ref{fig:TotalCrossSectionRatio} compares the new STAR total cross section results to CT14MC2nlo by plotting the ratio of STAR cross sections to the CT14MC2nlo cross sections for each boson. The error bars in the figure represent the total STAR measurement uncertainties and the CT14MC2nlo PDF uncertainties added in quadrature. The CT14MC2nlo PDF uncertainties used for $W^+$, $W^-$, and $Z$ cross sections were 5.9\%, 7.4\%, and 7.0\%, respectively.

\begin{table}[!h]
\caption{\label{tbl:Xstot} STAR total cross sections calculated from the combined 2011, 2012, and 2013 data sets. The columns labeled ``Stat.'' and ``Eff.'' represent the statistical uncertainty and the systematic uncertainty estimated from the efficiencies, respectively. The column ``Sys.'' includes all remaining systematic uncertainties, with the exception of the luminosity. The 9\% uncertainty associated with the luminosity measurement is not included in the table.}
\begin{ruledtabular}
\begin{tabular}{ccccc}
 & Cross Section (pb) & Stat. (pb) & Sys. (pb) & Eff. (pb) \\
\hline
$\sigma^{tot}_{W^+} \cdot B\left( W^+\rightarrow e^+\nu \right)$ & $143.0$ & $1.5$ & $2.5$ & $7.5$ \\
$\sigma^{tot}_{W^-} \cdot B\left( W^-\rightarrow e^- \bar{\nu} \right)$& $41.2$ & $1.0$ & $1.4$ & $2.3$  \\
$\sigma^{tot}_{Z} \cdot B\left( Z\rightarrow e^+e^- \right)$& $8.7$ & $0.5$ & $0.1$ & $0.9$  \\
\end{tabular}
\end{ruledtabular}
\end{table}

\section{Cross-Section Ratios}
\label{sec:ratios}
Equation~\ref{eq:fid} can also be used to compute the cross-section ratios $\sigma^{fid}_{W^+}/\sigma^{fid}_{W^-}$ and $\sigma^{fid}_{W}/\sigma^{fid}_{Z}$. A benefit to measuring the cross-section ratios rather than the absolute cross sections is that several systematic uncertainties are reduced or canceled. For example, the luminosity uncertainty in the cross-section ratios is canceled, while the tracking efficiency uncertainty is reduced in the $W/Z$ (5\%) measurement and canceled in the $W^+/W^-$ measurement.  
\subsection{$\mathbf{W}$ Cross-Section Ratio}\label{sec:RW}
The $W^+/W^-$ ratio is presented in eight pseudorapidity bins in the mid-pseudorapidity region ($|\eta_{e}|<1$), and in one intermediate pseudorapidity bin that covered $1.0  < \eta_{e} < 1.5$. This binning followed the same pseudorapidity binning used for the differential cross sections discussed in Sec.~\ref{sec:diffxsec}. The $W^+/W^-$ cross-section ratio was computed separately for each of the three data sets in the mid-pseudorapidity region, while the $W^+/W^-$ cross-section ratio in the intermediate pseudorapidity region covered by the EEMC was computed from the combined 2012 and 2013 data sets. 

\begin{center}
\begin{figure}[!th]
\center
\includegraphics[width = 1\columnwidth,keepaspectratio]{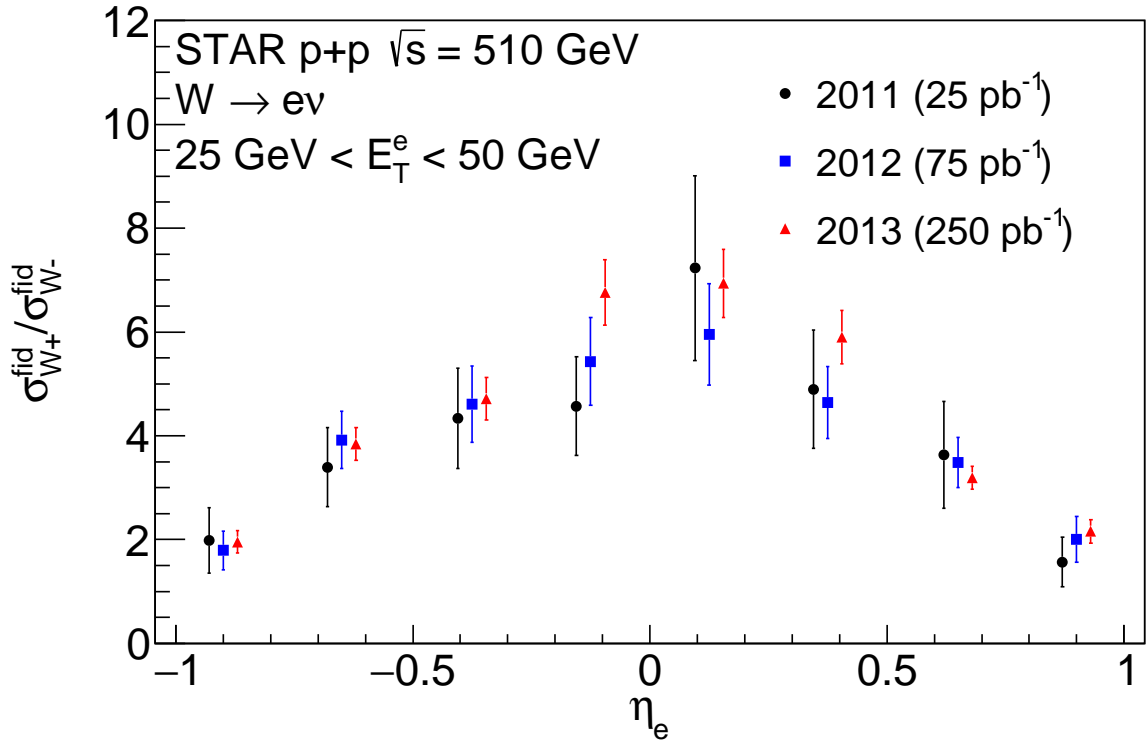}
\caption{\label{fig:RW-DataSets} Ratio of fiducial cross sections for production of $W^+$ and $W^-$ bosons plotted against the decay charged lepton pseudorapidity, $\eta_e$, for each of the three data sets: 2011 (black circle), 2012 (blue square), and 2013 (red triangle). For clarity, positions of the data points for the 2011, 2012, and 2013 data sets within each bin are offset by -0.03, 0.0, and 0.03. The error bars correspond to the statistical uncertainty associated with the cross-section ratio.}
\end{figure}
\end{center}

Figure~\ref{fig:RW-DataSets} shows a comparison of the $W^+/W^-$ cross-section ratios for each data set measured in the mid-pseudorapidity region as a function of pseudorapidity, where the error bars represent statistical uncertainties only. From the figure one can see consistency amongst the data sets and improvement in the statistical precision with each year. These values are plotted with an offset in $\eta_e$ for clarity.

\begin{center}
\begin{figure}[!h]
\includegraphics[width = 0.8\columnwidth,keepaspectratio]{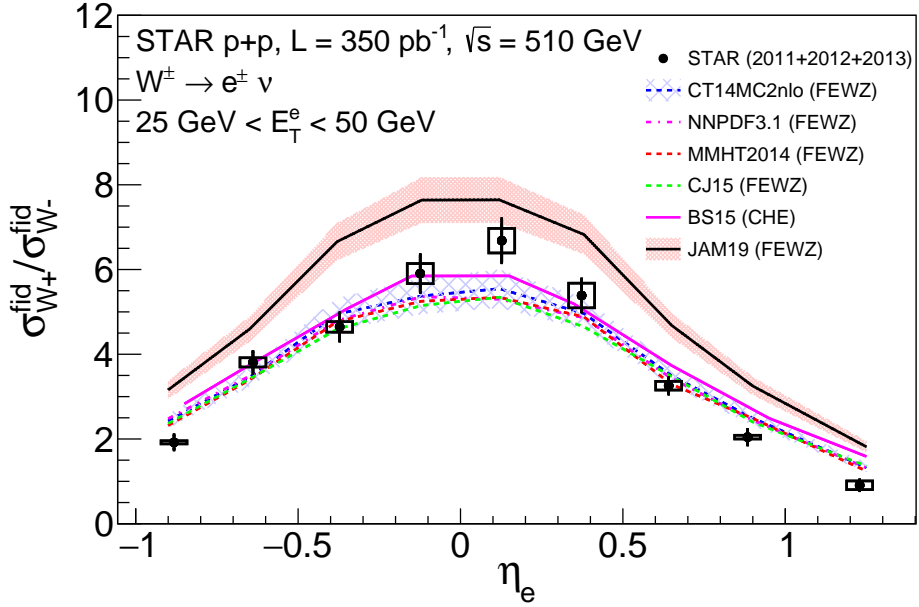}
\caption{\label{fig:RW} The combined (2011,2012, and 2013) results for the ratio of the fiducial cross sections for the production of $W^+$ and $W^-$ bosons plotted against the decay charged letpon pseudorapidity, $\eta_e$. The error bars represent the statistical uncertainty, whereas the rectangular boxes represent the systematic uncertainty for the respective data point. These measurements are compared to various theory predictions displayed in the legend.}
\end{figure}
\end{center}

Systematic uncertainties for the backgrounds were computed, as described in Sec.~\ref{sec:background}, for the pseudorapidity dependent $W^+$ and $W^-$ distributions. These uncertainties were then propagated to the $W^+/W^-$ cross-section ratios, which lead to about $\sim 2.5$\% ($4$\%) average uncertainty on the $W^+/W^-$ cross-section ratio measured in the mid- (intermediate) pseudorapidity regions. The efficiency uncertainties due to the energy scale, discussed in Sec.~\ref{sec:eff} were then propagated to the $W^+/W^-$ ratios measured in the mid- (intermediate) pseudorapidity region, which contributed $1.5$\% ($9$\%) to the total systematic uncertainty. An additional uncertainty that was studied is related to the difference in the $\eta_{e^+}$ and $\eta_{e^-}$ distributions in the intermediate pseudorapidity measurement. For measurements in the mid-pseudorapidity region these differences were negligible. However, in the intermediate pseudorapidity range the means of the two $\eta_e$ distributions differ by about 0.05. FEWZ was used to investigate how the $W^+/W^-$ cross-section ratio changes over this range using the CT14MC2nlo~\cite{Hou2017}, MMHT14~\cite{Harland-Lang2015}, and NNPDF3.1~\cite{Ball2017} NLO PDF sets. Based on this study, an uncertainty of 9\% was estimated and applied to the intermediate $W^+/W^-$ cross-section ratio. Figure~\ref{fig:RW} shows the $W^+/W^-$ cross-section ratios for the combined data sets plotted against the pseudorapidity. These measurements are also compared to NLO predictions using two theory frameworks (FEWZ~\cite{PhysRevD.86.094034} and CHE~\cite{PhysRevD.81.094020}), and various PDF inputs (CT14MC2nlo~\cite{Hou2017}, MMHT14~\cite{Harland-Lang2015}, BS15~\cite{BOURRELY2015307}, CJ15~\cite{Accardi:2016qay}, JAM19~\cite{Sato:2019yez}, and NNPDF 3.1~\cite{Ball2017}). The hatched uncertainty band represents the uncertainty associated with using the CT14MC2nlo PDF set within the FEWZ framework. The PDF sets are found to be consistent within the precision of the measured data. The results shown in Fig.~\ref{fig:RW} are listed in Table~\ref{tbl:RW}. 

\begin{table}[!h]
\caption{\label{tbl:RW} The combined (2011, 2012, and 2013) results for the ratio of the fiducial cross sections for production of $W^+$ and $W^-$ bosons in bins of the decay charged lepton pseudorapidity.}
\begin{ruledtabular}
\begin{tabular}{cccc}
$<\eta_{e}>$ & $\sigma^{fid}_{W^+}/\sigma^{fid}_{W^-}$ & Stat. & Sys.\\
\hline
$-0.88$ & $1.9$ & $0.2$ & $0.1$\\
$-0.64$ & $3.8$ & $0.3$ & $0.1$\\
$-0.37$ & $4.6$ & $0.3$ & $0.1$\\
$-0.12$ & $5.9$ & $0.4$ & $0.2$\\
$0.13$  & $6.7$ & $0.5$ & $0.3$\\
$0.37$  & $5.4$ & $0.4$ & $0.3$\\
$0.64$  & $3.3$ & $0.2$ & $0.1$\\
$0.88$  & $2.0$ & $0.2$ & $0.1$\\
$1.23$  & $0.9$ & $0.1$ & $0.1$\\
\end{tabular}
\end{ruledtabular}
\end{table}
%Figure~\ref{fig:RW} shows the $W^+/W^-$ cross-section ratios for the combined data sets plotted against the pseudorapidity. These measurements are also compared to NLO predictions using two theory frameworks (FEWZ~\cite{PhysRevD.86.094034} and CHE~\cite{PhysRevD.81.094020}), and various PDF inputs (CT14MC2nlo~\cite{Hou2017}, MMHT14~\cite{Harland-Lang2015}, BS15~\cite{BOURRELY2015307}, CJ15~\cite{Accardi:2016qay}, JAM19~\cite{Sato:2019yez}, and NNPDF 3.1~\cite{Ball2017}). The hatched uncertainty band represents the uncertainty associated with using the CT14 PDF set within the FEWZ framework. The PDF sets are found to be consistent within the precision of the measured data. The results shown in Fig.~\ref{fig:RW}, including the pseudorapidity details, are listed in Table~\ref{tbl:RW}. 
\subsection{$\mathbf{W}$ Cross-section Ratio PDF Impact}
\begin{center}
\begin{figure}[!h]
\includegraphics[width = 1\columnwidth,keepaspectratio]{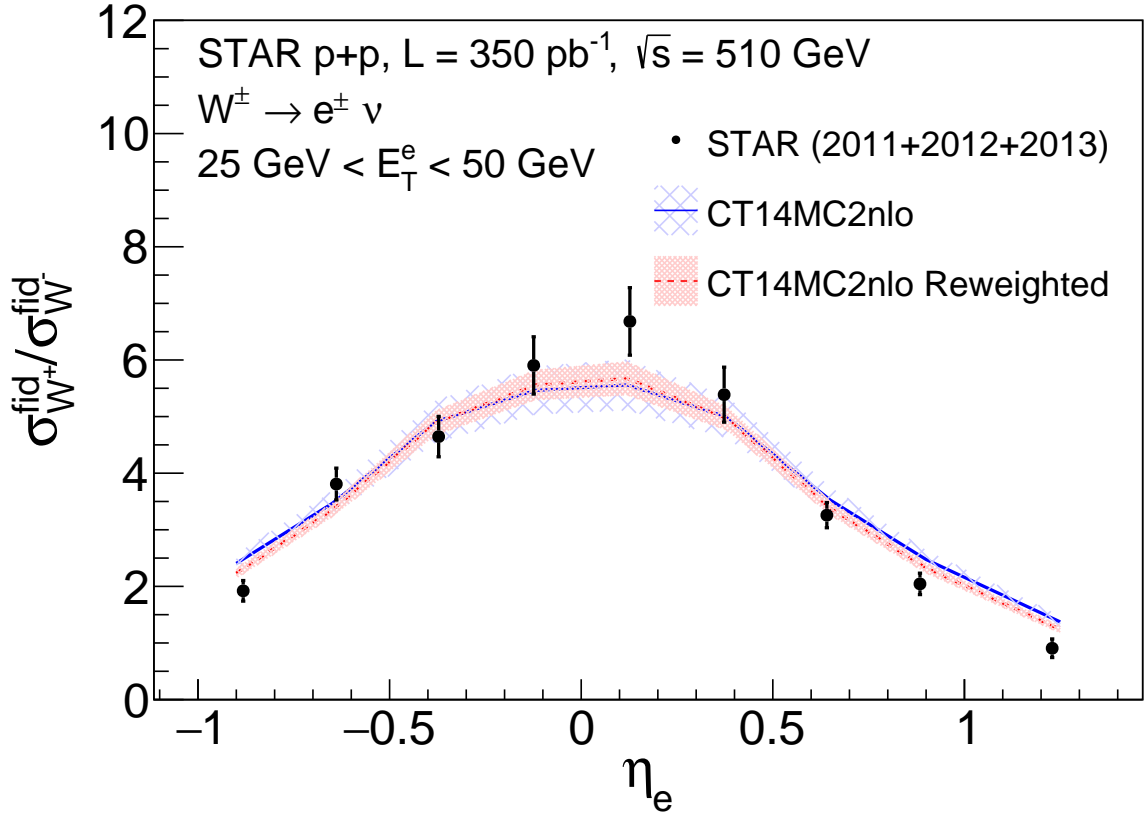}
\caption{\label{fig:RWfit} The combined results for the ratio of the fiducial cross sections for the production of $W^+$ and $W^-$ bosons compared to the predictions from the original and reweighted CT14MC2nlo PDF~\cite{Hou2017} predictions. The error bars on the STAR data represent the quadrature sum of the statistical and systematic uncertainties. The blue hatched band represents the CT14MC2nlo PDF uncertainty, while the red band shows the reweighted CT14MC2nlo PDF uncertainty after fitting the STAR data.}
\end{figure}
\end{center}

\begin{center}
\begin{figure}[!h]
\includegraphics[width = 0.8\columnwidth,keepaspectratio]{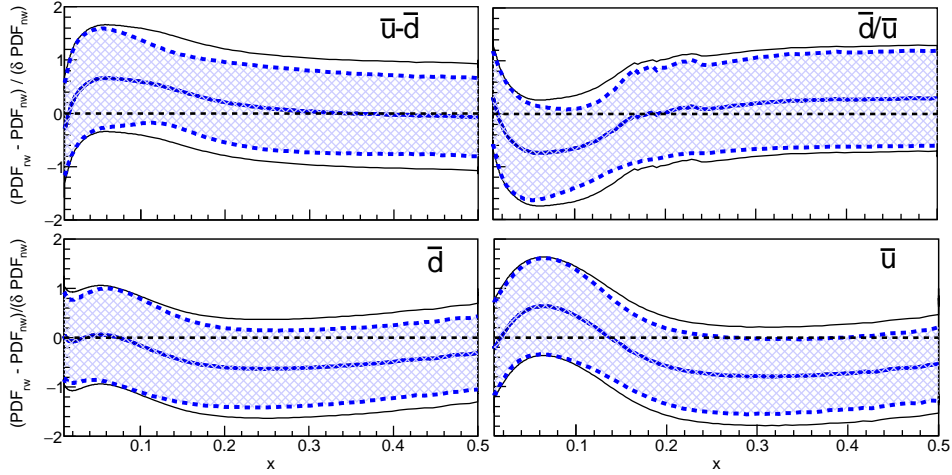}
\caption{\label{fig:PDFimpact}The impact of STAR $W^+/W^-$ data on $\bar{u} - \bar{d}$, $\bar{d}/\bar{u}$, $\bar{d}$, and $\bar{u}$ PDF distributions at $Q = 100$ GeV. The solid blue line shows the difference between the reweighted and nominal CT14MC2nlo PDF central value, normalized by the nominal PDF uncertainty. The hatched bands represent the ratio between the reweighted and nominal PDF uncertainties. The black lines represent the nominal PDF uncertainties from the solid blue line.}
\end{figure}
\end{center}

Ultimately, the results we presented are intended to be included in future global analyses to constrain PDF quark distributions. However, in the meantime we can assess the impact of these measurements through a PDF reweighting procedure. The $W^+/W^-$ cross-section ratio results discussed in Sec.~\ref{sec:RW} were used to reweight the CT14MC2nlo~\cite{Hou2017} PDF set using the procedure discussed in Ref.~\cite{BALL2011112,*BALL2011112e1,*BALL2011112e2}. FEWZ was used to evaluate the $W^\pm$ fiducial cross sections needed as input to evaluate the $W^+/W^-$ cross-section ratio for each of the 1000 CT14MC2nlo replicas. The result of this reweighting with the new STAR data is shown in Fig.~\ref{fig:RWfit} as a function of pseudorapidity. The red band is the reweighted distribution and the CT14MC2nlo uncertainties are given by the blue hatched band.~The impact of the STAR data on various PDF central distributions is assessed by investigating the difference between the reweighted PDF distribution ($PDF_{rw}$) and the nominal CT14MC2nlo PDF distribution ($PDF_{nw}$), normalized to the nominal PDF uncertainty ($\delta PDF_{nw}$).~Figure~\ref{fig:PDFimpact} shows the quantity $\left(PDF_{rw} - PDF_{nw}\right)/\left(\delta PDF_{nw}\right)$ (the blue solid line), plotted as a function of $x$ at the scale $Q$ = 100 GeV, for several PDF distributions ($\bar{u}-\bar{d}$, $\bar{d}/\bar{u}$, $\bar{d}$, and $\bar{u}$). The hatched bands in Fig.~\ref{fig:PDFimpact} represent the ratio between the reweighted and nominal PDF uncertainties, $\pm(\delta PDF_{rw}/\delta PDF_{nw})$, which are enclosed by blue dashed lines and can be used to assess the change in the PDF uncertainty. The black lines represent $\pm \delta PDF_{nw}$ uncertainties from the solid blue line. The difference between the solid black and dashed blue lines shows the change in uncertainty. On the other hand deviations of the solid blue line from zero represent changes in the central value of the nominal PDF set. From Fig.~\ref{fig:PDFimpact}, a clear but modest reduction in the uncertainty is seen in all of the distributions. Furthermore, all distributions show some modification to the nominal PDF's central values, which are generally within the one-sigma level. The change in the $\bar{d}/\bar{u}$ ratio is negative over the $x$ range of $0.05 - 0.2$, which indicates the reweighted PDF prefers to have a smaller central value of $\bar{d}/\bar{u}$ compared to the nominal PDF set. While at $x > 0.2$, the change is slightly positive indicating that the reweighted PDF prefers a larger $\bar{d}/\bar{u}$ than the nominal PDF.   

\subsection{$\mathbf{(W^+ + W^-)/Z}$ Cross-Section Ratio}\label{sec:WZ}
The $\sigma^{fid}_{W}/\sigma^{fid}_{Z}$ cross-section ratio was formed using Eq.~\ref{eq:fid} and adding the $W^+$ and $W^-$ fiducial cross sections in the central pseudorapidity region ($|\eta_e|<1$). The systematic uncertainties for the $W$ cross sections were evaluated as discussed in Sec.~\ref{sec:RW}, with the exception of the track reconstruction uncertainty, and were propagated to the $(W^+ + W^-)/Z$ cross-section ratio measurement. The systematic uncertainty associated with the track reconstruction efficiency was estimated at $5$\% due to partial cancellation. 

\begin{center}
\begin{figure}[!h]
\includegraphics[width = 0.8\columnwidth,keepaspectratio]{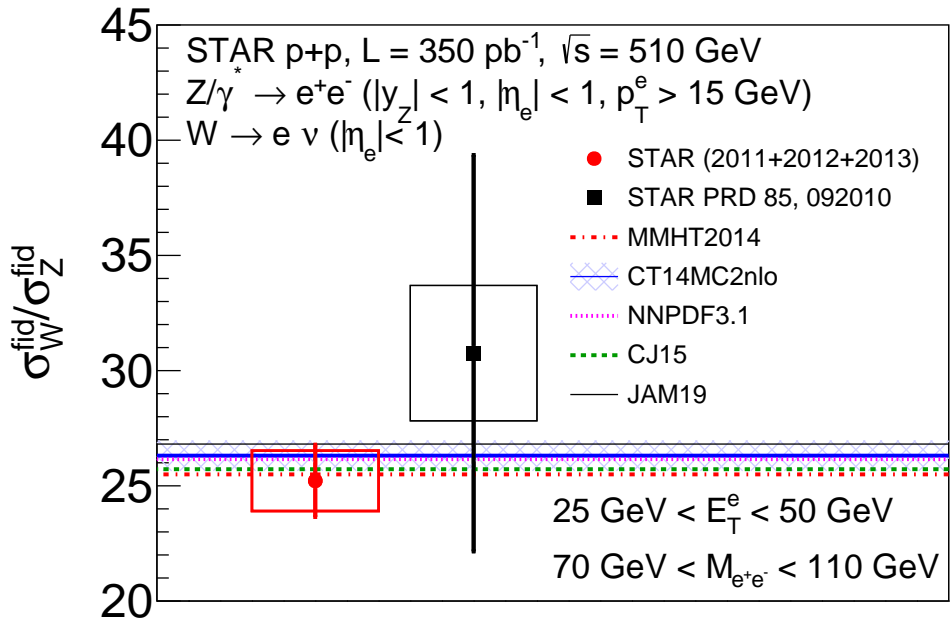}
\caption{\label{fig:WZ}The measured $(W^+ + W^-)/Z$ (red circle marker) for the combined data sets (2011-2013). Compared to this measurement is the $(W^+ + W^-)/Z$ computed from the STAR 2009 data set~\cite{STAR:2011aa}(black square marker), and evaluations using the FEWZ framework~\cite{PhysRevD.86.094034} and several input NLO PDF sets (MMHT14~\cite{Harland-Lang2015}, CT14MC2nlo~\cite{Hou2017}, NNPDF3.1~\cite{Ball2017}, CJ15~\cite{Accardi:2016qay}, and JAM19~\cite{Sato:2019yez}).}
\end{figure}
\end{center}

The measured $\sigma^{fid}_{W}/\sigma^{fid}_{Z}$ cross-section ratio for the combined 2011, 2012 and 2013 data sets was found to be $25.2\pm 1.6_{(\text{stat.})} \pm 1.3_{(\text{syst.})}$, and is shown in Fig.~\ref{fig:WZ}. The $(W^+ + W^-)/Z$ cross-section ratio is compared to NLO evaluations using the FEWZ framework and several input PDF sets. This measurement is consistent with the FEWZ predictions for all PDF sets investigated and will allow us to further constrain the sea quark PDFs.~The uncertainty associated with the $W/Z$ cross-section ratio calculated from CT14MC2nlo replicas was estimated to be 2.5\% (blue hatched band), based on the distribution's RMS. Also included is the $(W^+ + W^-)/Z$ cross-section ratio computed from the $W$ and $Z$ fiducial cross sections from the 2009 STAR $p+p$ data set~\cite{STAR:2011aa}. The error bars represent the statistical uncertainties, while the boxes represent the total systematic uncertainties.\\

\section{Summary}
\label{sec:summary} 
STAR has measured the $W$ and $Z$ total and differential cross sections, along with the $W^+/W^-$ and $(W^+ + W^-)/Z$ cross-section ratios in $p+p$ collisions at center of mass energies of $500$ GeV and $510$ GeV at RHIC, using the total luminosity of 350 pb$^{-1}$. These measurements not only provide additional high $Q^2$ data to be used in future global analyses to help constrain PDFs, but also serve as complementary measurements to other experiments. In particular, our total and differential $W$ and $Z$ cross sections along with the $(W^+ + W^-)/Z$ cross-section ratio, will complement LHC's $W$ and $Z$ production program by providing data at lower $\sqrt{s}$ and sensitivity at larger $x$. Our $W^+/W^-$ cross-section ratio measurement, which is particularly sensitive to the $\bar{d}/\bar{u}$ sea quark distribution~\cite{Nadolsky:Private} (Eq.~\ref{eq:RW-Theory}), provides an alternative method to study the $\bar{d}/\bar{u}$ distribution which is complementary to the measurements performed by the NuSea and SeaQuest experiments. 

Using our pseudorapidity dependent $W^+/W^-$ cross-section ratio results in a PDF reweighting study, we find sensitivity to the sea quark distributions. Our study shows modest improvement in the uncertainties of several distributions, in particular the $\bar{d}/\bar{u}$ and $\bar{u}-\bar{d}$ distributions, as well as a change in the central values. 

Overall we find good agreement between our measurements and the current PDF distributions. Inclusion of these data into future global fits will help to constrain the PDFs.
\section{Acknowledgments}
\label{sec:ackno}
We thank the RHIC Operations Group and RCF at BNL, the NERSC Center at LBNL, and the Open Science Grid consortium for providing resources and support.  This work was supported in part by the Office of Nuclear Physics within the U.S. DOE Office of Science, the U.S. National Science Foundation, the Ministry of Education and Science of the Russian Federation, National Natural Science Foundation of China, Chinese Academy of Science, the Ministry of Science and Technology of China and the Chinese Ministry of Education, the Higher Education Sprout Project by Ministry of Education at NCKU, the National Research Foundation of Korea, Czech Science Foundation and Ministry of Education, Youth and Sports of the Czech Republic, Hungarian National Research, Development and Innovation Office, New National Excellency Programme of the Hungarian Ministry of Human Capacities, Department of Atomic Energy and Department of Science and Technology of the Government of India, the National Science Centre of Poland, the Ministry  of Science, Education and Sports of the Republic of Croatia, RosAtom of Russia and German Bundesministerium fur Bildung, Wissenschaft, Forschung and Technologie (BMBF), Helmholtz Association, Ministry of Education, Culture, Sports, Science, and Technology (MEXT) and Japan Society for the Promotion of Science (JSPS).
\bibliographystyle{apsrev4-1}
\bibliography{RW2019}

\end{document}